 \documentclass[5p,times,sort&compress]{elsarticle}          % two-column typeset (journal preview)
\usepackage[utf8]{inputenc}
\usepackage{amsmath}
\usepackage{amssymb}
\usepackage{amsopn}
\usepackage{mathtools}
\usepackage{bm}
\usepackage{textcomp}
\usepackage{gensymb}
\usepackage{chemformula}
\usepackage{siunitx}
\DeclareSIUnit{\atpercent}{at.\%}
\DeclareSIUnit{\angstrom}{\protect \text {Å}}
\usepackage{graphicx}
\graphicspath{{./figs/}}
\usepackage[table,dvipsnames]{xcolor}
\definecolor{HeaderLine}{RGB}{0, 131, 162}
\usepackage{booktabs}
\usepackage{multirow}
\usepackage{array}
\usepackage{tabularx}
\usepackage{float}
\usepackage{subcaption}
\usepackage{geometry}
\usepackage[ruled,vlined]{algorithm2e}
\usepackage{dirtytalk}
\usepackage{cuted}
\usepackage[
  breaklinks=true,
  plainpages=false,
  hypertexnames=false,
  pdfpagelabels=true
]{hyperref}
\hypersetup{hidelinks}
\usepackage{cleveref}
\usepackage{xcolor}
\usepackage{overpic}
\newcommand{\etal}{\textit{et al.~}}

\makeatletter
\def\ps@pprintTitle{%
 \let\@oddhead\@empty
 \let\@evenhead\@empty
 \def\@oddfoot{}%
 \let\@evenfoot\@oddfoot}
\makeatother

\begin{document}

\sloppy

\begin{frontmatter}

\title{From MLIPs to Microstructure: A  High-Throughput Computational Framework to Design Spinodal Alloys in High-Dimensional Composition Spaces via Analytic Derivatives of CALPHAD Model Predictions}

%\title{From MLIPs to CALPHAD to Microstructure: A  High-Throughput Computational Framework to Design Spinodal Alloys in High-Dimensional Composition Spaces}

\author[1]{Courtney Kunselman\corref{cor1}}
\ead{cjkunselman18@tamu.edu}
\author[1]{Do\u{g}uhan Sar\i{}t\"urk}
\author[1]{Siya Zhu}
\author[1]{Vahid Attari}
\author[1,2,3]{Raymundo Arr\'{o}yave}

\cortext[cor1]{Corresponding author}

\affiliation[1]{organization={Department of Materials Science and Engineering, Texas A\&M University},
                city={College Station},
                postcode={77843}, 
                state={TX},
                country={USA}}

\affiliation[2]{organization={J. Mike Walker '66 Department of Mechanical Engineering, Texas A\&M University},
                city={College Station},
                postcode={77843}, 
                state={TX},
                country={USA}}

\affiliation[3]{organization={Wm Michael Barnes '64 Department of Industrial and Systems Engineering, Texas A\&M University},
                city={College Station},
                postcode={77843}, 
                state={TX},
                country={USA}}

\begin{abstract}    
   Identifying regions of design space subject to spinodal decomposition is a critical component of alloy design in high-dimensional composition spaces. In cases where designers are seeking to exploit spinodal microstructures to tailor alloy properties, prediction of microstructure evolution and morphology is also needed. In this work, we present a Machine Learning Interatomic Potential (MLIP)-trained, CALPHAD-based, open-source workflow for high-throughput microstructure stability analysis and visualization. In this workflow, coherent strain contributions are captured via high-throughput MLIP elastic constant calculations. To predict microstructure morphology for compositions of interest, MLIP-generated thermodynamic models are fed into an elasto-chemical phase field simulation. Both stability analyses and phase-field simulations utilize analytically-derived Gibbs energy Hessians to improve computational efficiency and accuracy over finite difference approximations. We demonstrate this workflow by investigating microstructure stability in the Hf-Nb-Ti-V quaternary system.
   
\end{abstract}


\end{frontmatter}

%\linenumbers

\section{Introduction}

Spinodal decomposition is a barrierless phase separation process driven by uphill diffusion~\cite{cahn1961spinodal}, producing composition-modulated microstructures whose morphology depends on composition, elastic anisotropy, and stage of evolution. Because microstructure sets properties, alloy design requires identifying the composition--temperature regions where a phase is unstable to fluctuations. In some cases, spinodal decomposition is undesirable, causing hardening and embrittlement of otherwise ductile single-phase alloys \cite{wang2023spinodal} or opening a pathway to brittle secondary phases \cite{dasari2023crystallographic}. On the other hand, spinodal and spinodal-assisted microstructures have recently been exploited to overcome the strength-ductility tradeoff \cite{an2021spinodal,lee2025spinodal,zhang2022refinement}, tailor magnetic properties \cite{rao2021beyond,dong2024cooperative}, and improve thermoelectric performance \cite{duan2025spinodal}. In equiatomic HfNbTiV, for instance, An \etal reported coherency-induced lattice strains of up to \SI{5}{\percent} in the minority BCC phase---among the largest reported in refractory high entropy alloys (HEAs)~\cite{an2021spinodal}. 

Regardless of whether the designer seeks to exploit or avoid spinodal decomposition, that region must first be located within a design space that is often high-dimensional. Because exhaustive experimental characterization of these large design spaces is infeasible, computational models that predict microstructure stability are essential for identifying spinodal regions. Recent work to this end includes high-throughput \textit{ab initio} calculations coupled with cluster expansion to construct CALPHAD-style Gibbs energy models \cite{usanmaz2016first}, and a procedure that requires no functional form of the free energy, modeling spinodal decomposition as a nucleation event infinitesimally close to the spinodal boundary \cite{divilov2024priori}. While these approaches require no empirical data and have been well-validated against experiment, they were demonstrated only on binary systems. For higher-dimensional composition spaces, Koneru, Kadirvel, and Wang recently developed a CALPHAD-based high-throughput framework for the multi-principal element alloy (MPEA) or HEA design problem \cite{koneru2022high,kadirvel2022exploration}. This framework, now implemented in Pandat \cite{cao2009pandat}, provides gauge-invariant prediction and visualization of Gibbs energy surface geometry, including eigenvalues and eigenvectors of the Gibbs energy Hessian, spinodal temperatures, and initial composition fluctuations. 

In this work, we take the framework of Koneru, Kadirvel, and Wang as a starting point and address four of its limitations. First, its numerical evaluation of Gibbs energy Hessians becomes computationally expensive as the number of components grows---we use the recently formalized Jansson derivative method to calculate analytic Hessians instead \cite{kunselman2024analytically}. Second, it relies on commercial CALPHAD databases that are unavailable for many refractory HEA systems---we use the recently developed PhaseForge package \cite{sariturk_2025_15730911} to construct multicomponent CALPHAD models from machine learning interatomic potentials (MLIP). Third, and most consequential for real alloys, its chemical driving force does not capture the coherency strains that shift spinodal boundaries in refractory HEAs---we add an estimate of the elastic energy contribution following Cahn \cite{cahn1962coherent}, which penalizes phase separation and shrinks the miscibility gap and spinodal region. Fourth, its constant-core-component pseudoternary diagrams become unwieldy above four components---we use affine projections \cite{vela2025visualizing} to summarize high-dimensional data on a single two-dimensional plot. 

The entire workflow is packaged in an open-source tool available through the MaterialsFramework interface \cite{sariturk_2025_15731044}. Through this framework, modelers have whitebox access to the CALPHAD model, and as demonstrated in \cite{kunselman2025construction}, they can further refine parameters against any available computational or experimental data using ESPEI \cite{bocklund2019espei}. The remainder of this work presents the theory and methodology of determining microstructure stability from the Gibbs energy description of a phase, and then applies the framework to investigate the stability of the BCC phase in the \ch{Hf-Nb-Ti-V} system. Beyond spinodal prediction, we also introduce an elasto-chemical phase field approach that uses Jansson-derivative surrogates for the Gibbs energy Hessian to predict microstructure evolution. The predicted microstructures are compared against the experimental observations of An~\etal~\cite{an2021spinodal} to assess where the pipeline succeeds and where it falls short.

\section{Theory}
\subsection{Developing Gibbs Energy Models From Universal Potentials}
\label{sec:gibbs_models}

The stability analysis described in the following section requires a molar Gibbs energy description of the solid solution phase of interest as a function of composition and temperature. We use the PhaseForge package \cite{sariturk_2025_15730911} to construct CALPHAD-compatible Gibbs energy models from MLIPs as described in \cite{zhu2025accelerating,zhu2025machine}, replacing computationally expensive density functional theory (DFT) calculations with MLIP evaluations that are several orders of magnitude faster while maintaining thermodynamic accuracy.

The approach is built on the use of special quasirandom structures (SQS) \cite{zunger1990special} to represent disordered solid solutions in a periodic cell. An SQS is a periodic supercell whose short-range order parameters match those of a perfectly random alloy up to a pre-determined neighbor range, enabling the disordered state to be represented without requiring large supercells.

Following the compound energy formalism (CEF) as implemented in the Alloy Theoretic Automated Toolkit (ATAT) \cite{van2002alloy}, the molar Gibbs energy of solution phase $\beta$ with site fractions $\boldsymbol{y}$ at temperature $T$ is given by
\begin{equation}\label{eq:gibbs_cef}
\begin{split}
    G^\beta_m(\boldsymbol{y}, T) =& \left(G^\beta_{\mathrm{calc,nc}}(\boldsymbol{y}, T) - \sum_i x_i\, G^{\alpha_i}_{\mathrm{calc,nc}}(T)\right) \\&+ \sum_i x_i\, G^{\alpha_i}_{\mathrm{SGTE}}(T) - T S_{\mathrm{id}}(\boldsymbol{y}),
\end{split}
\end{equation}
where $G^\beta_{\mathrm{calc,nc}}$ is the Gibbs energy of phase $\beta$ computed from MLIP total energies without any configurational entropy contribution, $G^{\alpha_i}_{\mathrm{calc,nc}}$ is the corresponding quantity for pure component $i$ in its reference crystal structure $\alpha_i$, $G^{\alpha_i}_{\mathrm{SGTE}}$ is the unary Gibbs energy of component $i$ taken from the Scientific Group Thermodata Europe (SGTE) database \cite{dinsdale1991sgte}, $x_i$ is the mole fraction component $i$, and $S_{\mathrm{id}}(\boldsymbol{y})$ is the ideal configurational entropy. For a single-sublattice solution phase, the ideal configurational entropy reduces to
\begin{equation}\label{eq:conf_entropy}
    S_{\mathrm{id}}(\boldsymbol{y}) = -R\sum_i x_i\ln x_i,
\end{equation}
where $R$ is the ideal gas constant.

For each SQS supercell, the quantity $G^\beta_{\mathrm{calc,nc}}$ at \SI{0}{\kelvin} is taken directly from the MLIP-relaxed potential energy per atom. Structural relaxations are performed through the MaterialsFramework calculator interface \cite{sariturk_2025_15731044} with a force convergence criterion of $f_{\mathrm{max}} < \SI{0.001}{\electronvolt/\angstrom}$. For the BCC phase in the \ch{Hf-Nb-Ti-V} system, the compositional dependence of the mixing energy is captured sufficiently by static formation energies, and thermal contributions are accounted for through the SGTE unary data.

All SQS structures are evaluated using ORB v2~\cite{neumann2024orb}, a non-equivariant graph-network-based universal MLIP trained on the MPtrj~\cite{deng2023chgnet} and Alexandria~\cite{schmidt2023alexandria} datasets. ORB v2 was selected because it produced the lowest formation-energy errors among five universal MLIPs benchmarked for CALPHAD-based calculations~\cite{zhu2025accelerating}. Once formation energies have been computed for SQS supercells spanning the composition space of each phase, the composition-dependent mixing contribution embedded in $G^\beta_{\mathrm{calc,nc}}$ is parameterized using a Redlich-Kister polynomial expansion. For a binary \ch{A-B} subsystem, the excess Gibbs energy is expressed as
\begin{equation}\label{eq:rk}
    G^\beta_{m,\mathrm{xs}} = x_A x_B \sum_{\nu=0}^{k} L^{(\nu)}_{A,B}\,(x_A - x_B)^\nu,
\end{equation}
where $L^{(\nu)}_{A,B}$ are the $\nu$-th order binary interaction parameters determined by least-squares fitting to the MLIP-computed formation energies. Higher-order ternary interactions are incorporated analogously and extracted by fitting to SQS energies at compositions spanning the multicomponent space.

The fitted interaction parameters are assembled into a thermodynamic database (TDB) file compatible with PyCalphad \cite{otis2017pycalphad}. The resulting TDB for the BCC phase in the \ch{Hf-Nb-Ti-V} system serves as the thermodynamic input for the stability analysis described below. Although pure Hf and Ti adopt the HCP structure at room temperature, the BCC\textsubscript{A2} phase is stabilized at the concentrated compositions studied here and is the experimentally observed structure~\cite{an2021spinodal}. The analysis throughout this work is therefore restricted to the BCC\textsubscript{A2} phase.

\subsection{Determining Stability From the Gibbs Energy Hessian}
\label{sec:stability_theory}

Gibbs first introduced the concept of a thermodynamic stability limit, or spinodal, by considering composition fluctuations leading to continuous changes in fluid phases \cite{gibbs1878equilibrium}. He determined that the limit of stability for such a phase relative to these fluctuations occurs where the derivatives of the chemical potential of each component with respect to the density of its corresponding component are equal to zero. That is, a system stable to all composition fluctuations experiences positive changes in all of its chemical potentials when the content of the corresponding component is increased. As Cahn showed \cite{cahn1961spinodal}, for binary system $A-B$ with $B$ as the dependent component, this limit of stability simplifies to
\begin{equation}\label{second_der_G}
    \frac{d^2G^\alpha_m}{dx^2_A}=0,
\end{equation}
\noindent where $G^\alpha_m$ is the molar Gibbs energy for phase $\alpha$ and $x_A$ is the composition of component $A$. 

Following Gibbs and Cahn, the work of de Fontaine derived the conditions for stability for solid phases in multicomponent systems by including coherency stress contributions to the free energy in both continuum \cite{de1967computer} and microscopic \cite{de1972analysis} contexts. In recent works, Kadirvel \etal provide an elegant condensed proof of these conditions using the continuum formulation \cite{kadirvel2022exploration}, and Morral and Chen apply de Fontaine's findings to comprehensively extend Gibbs' work to the solid phase, multicomponent system context. We summarize their combined insight below.

Consider solid solution phase $\alpha$, an $n$-component phase with composition $\boldsymbol{x}$ for which we define the $n$-th component as the dependent composition variable. Let $\boldsymbol{\Delta}$ be a local arbitrary composition fluctuation in phase $\alpha$ defined for all independent composition variables. To maintain mass balance, a local composition fluctuation of $-\boldsymbol{\Delta}$ must also occur for an equivalent amount of the system. The molar Gibbs energy change of the portion of phase $\alpha$ affected by this composition fluctuation is given by
\begin{equation}\label{eq:energy_change}
\Delta G^\alpha_m = \frac{1}{2}G^\alpha_m(\boldsymbol{x}+\boldsymbol{\Delta}) + \frac{1}{2}G^\alpha_m(\boldsymbol{x}-\boldsymbol{\Delta}) -G^\alpha_m(\boldsymbol{x}).
\end{equation}
\noindent Using a second order Taylor series expanded around $\boldsymbol{x}$, we can approximate $G^\alpha_m(\boldsymbol{x}+\boldsymbol{\Delta})$ as 
\begin{equation}\label{eq:Taylor_1}
    G^\alpha_m(\boldsymbol{x}+\boldsymbol{\Delta})\approx G^\alpha_m(\boldsymbol{x}) + \sum_{i=1}^{n-1}\frac{\partial G^\alpha_m}{\partial x_i}\Delta_i+\frac{1}{2}\sum_{i=1}^{n-1}\sum_{j=1}^{n-1}\frac{\partial^2G^\alpha_m}{\partial x_i\partial x_j}\Delta_i\Delta_j
\end{equation}
\noindent and $G^\alpha_m(\boldsymbol{x}-\boldsymbol{\Delta})$ as
\begin{equation} \label{eq:Taylor_2}
    G^\alpha_m(\boldsymbol{x}-\boldsymbol{\Delta})\approx G^\alpha_m(\boldsymbol{x}) - \sum_{i=1}^{n-1}\frac{\partial G^\alpha_m}{\partial x_i}\Delta_i+\frac{1}{2}\sum_{i=1}^{n-1}\sum_{j=1}^{n-1}\frac{\partial^2G^\alpha_m}{\partial x_i\partial x_j}\Delta_i\Delta_j
\end{equation}
\noindent where all derivatives are evaluated at $\boldsymbol{x}$.

Substituting \Cref{eq:Taylor_1,eq:Taylor_2} into \Cref{eq:energy_change} gives
\begin{equation} \label{eq:Hessian}
\Delta G^\alpha_m = \sum_{i=1}^{n-1}\sum_{j=1}^{n-1}\frac{\partial^2G^\alpha_m}{\partial x_i\partial x_j}\Delta_i\Delta_j=\boldsymbol{\Delta}^TH\boldsymbol{\Delta}
\end{equation}
\noindent where $H$ is the Hessian of $G^\alpha_m$ with respect to the independent composition variables. Assuming that all entries of the Hessian are continuous throughout the composition domain, we can claim that $H$ is a symmetric matrix with real eigenvalues $\lambda_1,\hdots,\lambda_{n-1}$ and a corresponding set of eigenvectors $\boldsymbol{v}_1,\hdots,\boldsymbol{v}_{n-1}$ which form an orthonormal basis in the space defined by the independent composition variables. Thus, $\boldsymbol{\Delta}$ can be written as a linear combination of this basis:
\begin{equation} \label{eq:projection}
    \boldsymbol{\Delta} = \sum_{i=1}^{n-1} (\boldsymbol{\Delta}^T\boldsymbol{v}_i)\boldsymbol{v}_i.
\end{equation}
\noindent Substituting \Cref{eq:projection} into \Cref{eq:Hessian} and using the relationship between eigenvalues and eigenvectors ($H\boldsymbol{v}_i=\lambda_i\boldsymbol{v}_i$) gives
\begin{equation}\label{eq:final_delta_G}
\begin{split}
    \Delta G^\alpha_m &= \left(\sum_{i=1}^{n-1}(\boldsymbol{\Delta}^T\boldsymbol{v}_i)\boldsymbol{v}_i\right)^TH\left(\sum_{i=1}^{n-1}(\boldsymbol{\Delta}^T\boldsymbol{v}_i)\boldsymbol{v}_i\right)\\&=\left(\sum_{i=1}^{n-1}(\boldsymbol{\Delta}^T\boldsymbol{v}_i)\boldsymbol{v}_i\right)^T\left(\sum_{i=1}^{n-1}(\boldsymbol{\Delta}^T\boldsymbol{v}_i)\lambda_i\boldsymbol{v}_i\right)=\sum_{i=1}^{n-1}\lambda_i(\boldsymbol{\Delta}^T\boldsymbol{v}_i)^2.
    \end{split}
\end{equation}
\noindent Thus, as \Cref{eq:final_delta_G} illustrates, such a composition fluctuation can only lower the Gibbs energy if at least one of the eigenvalues is negative, and a zero-valued eigenvalue represents the limit of stability, or the spinodal. If only one eigenvalue, say $\lambda_1$, is negative, the system is conditionally unstable to composition fluctuations in the direction of $\boldsymbol{v}_1$, or equivalently, to fluctuations with large projection $\boldsymbol{\Delta}^T\boldsymbol{v}_1$. More negative eigenvalues lead to more directions for $\boldsymbol{\Delta}$ to be unstable, and $n-1$ negative eigenvalues imply that the solid solution is unstable to any $\boldsymbol{\Delta}$.

As Morral and Chen \cite{morral2021stability} and Kadirvel \etal \cite{kadirvel2022exploration} both report, these eigenvalues and eigenvectors are not gauge invariant. That is, $\lambda_1,\hdots,\lambda_{n-1}$ and $\boldsymbol{v}_1,\hdots,\boldsymbol{v}_{n-1}$ depend on the choice of the dependent component. This is due to a combination of the coordinate axes being non-orthogonal and the change in orientation of these axes with the variation of the dependent component. 

To address this issue, Kadirvel \etal \cite{kadirvel2022exploration} followed the approach of Singh \etal \cite{singh2015atomic} which involves a change of basis operation transforming $H$ from Gibbs simplex space to Cartesian space, performing the eigenanalysis in Cartesian space, and finally transforming the resulting eigenvectors back to Gibbs space (the eigenvalues calculated in Cartesian space remain invariant under this final transformation). Singh \etal \cite{singh2015atomic} provide an iterative procedure for constructing such an $N$-dimensional linear transformation, and Kadirvel \etal \cite{kadirvel2022exploration} explicitly provide this matrix for systems with ten components or fewer. In this work, we adopt this approach to calculate and visually display the distribution of gauge-invariant eigenvalues. 

Finally, for arbitrarily-complex phases modeled with internal degrees of freedom (or equivalently, for any phase with a Gibbs energy function that is multivalued in composition space), calculating the Gibbs energy Hessian requires differentiating the outputs of single-phase energy minimization calculations. These equilibrium calculations prescribe the composition of the phase (not the system) and determine the local minimum solution (which may be metastable) under this composition constraint. For phase $\alpha$ in ternary system A-B-C, it is well known that the first derivative of $G^\alpha_m$ with respect to $x_A$ can be written as a simple linear combination of chemical potentials
\begin{equation}\label{eq:first_derivative}
\frac{\partial G^\alpha_m}{\partial x_A}=\mu_A-\mu_C
\end{equation}
\noindent where $C$ is the dependent component. Thus, second order derivatives can be expressed as 
\begin{equation}\label{eq:second_derivatives}
    \frac{\partial^2G^\alpha_m}{\partial x_B\partial x_A} = \frac{\partial \mu_A}{\partial x_B} - \frac{\partial \mu_C}{\partial x_B},
\end{equation}
\noindent but calculation of \Cref{eq:second_derivatives} is less trivial. In this work, we use the recently formalized Jansson derivative method \cite{kunselman2024analytically}, an adjoint-like technique for calculating analytic derivatives of thermodynamic properties at equilibrium with respect to CALPHAD model equilibrium calculations, to efficiently calculate the derivatives of the chemical potentials in \Cref{eq:second_derivatives} rather than relying on numerical approximations. To mitigate numerical ill-conditioning which can lead to convergence failures, a small offset of $1\times10^{-4}$ was added to any zero components of the composition vector and subtracted from the largest non-zero component. All Gibbs energy Hessians are computed using PyCalphad \cite{otis2017pycalphad} (development version commit a8de09f).

Using the procedure described above, we can use Jansson derivatives to calculate the chemical Hessian and corresponding chemical spinodal from a CALPHAD Gibbs energy description of a solution phase. However, in many solid solutions, composition fluctuations can lead to substantial elastic energy contributions due to coherency strains. These coherency strains penalize composition fluctuations and when their elastic energy contributions are added to the chemical Gibbs energy, the spinodal region shrinks. As Cahn shows in \cite{cahn1962coherent}, when coherency strains are considered for an infinite binary anisotropic solid with cubic symmetry, the limit of stability relative to an arbitrary infinitesimal composition fluctuation given in \Cref{second_der_G} becomes
\begin{equation}
    \frac{d^2G^\alpha_m}{dx^2_A}+2\eta_A^2YV_m=0
\end{equation}
\noindent where $V_m$ is the molar volume, 
\begin{equation}
    Y=
    \begin{cases}
        \frac{(C_{11}+2C_{12})(C_{11}-C_{12})}{C_{11}} & \text{if }\hspace{.1 in} 2C_{44}-C_{11}+C_{12}>0\\
        \frac{6(C_{11}+2C_{12})C_{44}}{4C_{44}+C_{11}+2C_{12}} & \text{if } \hspace{.1 in} 2C_{44}-C_{11}+C_{12}<0,
    \end{cases}
\end{equation}
\noindent where $C_{11}$, $C_{12}$, and $C_{44}$ are the independent elastic constants for a cubic system, and $\eta_A$ is the linear strain per unit difference in component $A$:
\begin{equation}
\eta_A=\frac{1}{a}\frac{\partial a}{\partial x_A}
\end{equation}
\noindent where $a$ is the lattice parameter. Extending to the multicomponent context \cite{de1973analysis}, entries of the coherent Hessian $H^{coh}$ are given by
\begin{equation}
    H^{coh}_{ab}=\frac{\partial^2G^\alpha_m}{\partial x_B\partial x_A}+2\eta_A\eta_BYV_m
\end{equation}
\noindent where $\frac{\partial^2G^\alpha_m}{\partial x_B\partial x_A}$ is the chemical contribution as calculated in \Cref{eq:second_derivatives}.

In this work, we use the bond-based lattice parameter prediction model developed by Tandoc \etal \cite{tandoc2025bond} to calculate $\boldsymbol{\eta}$ and $V_m$. Under this model, the average lattice parameter $\bar{a}$ for a solid solution BCC phase is given by
\begin{equation}\label{lattice_param}
    \bar{a}=\frac{2\bar{d}_{111}}{\sqrt{3}}
\end{equation}
\noindent where $\bar{d}_{111}$ is the average bond length along the $\langle 111 \rangle$ close-packed direction. The average bond length is calculated through
\begin{equation}
    \bar{d}_{111}=\sum_i\sum_jx_ix_jd^{ij}_{111}
\end{equation}
\noindent where $d^{ij}_{111}$ is the first nearest neighbor (FNN) bond length for an $i-j$ pair along the $\langle 111 \rangle$ close-packed direction, following the framework of Tandoc \etal \cite{tandoc2025bond}. Same-species bond lengths $d^{ii}_{111}$ are extracted from MLIP-relaxed pure BCC cells via $d^{ii}_{111} = a_i^{\text{BCC}}\sqrt{3}/2$, while cross-species lengths $d^{ij}_{111}$ ($i \neq j$) are obtained from MLIP-relaxed B2 (\ch{CsCl}-type) binary ordered cells via $d^{ij}_{111} = a_{ij}^{\text{B2}}\sqrt{3}/2$. All relaxations for bond length extraction are performed using the \texttt{BondLatticeParameter} tool within MaterialsFramework \cite{sariturk_2025_15731044} with the ORB v2 universal MLIP~\cite{neumann2024orb}. Using this model for BCC structures, the molar volume is then given by
\begin{equation}
V_m=\frac{\bar{a}^3N_A}{2}
\end{equation}
\noindent where $N_A$ is Avogadro's number. To calculate derivatives of the lattice parameter for $\eta$, a dependent component must be assigned which must remain consistent throughout the calculation of the coherent Hessian. Let $k$ be the index of the dependent component. Then
\begin{equation}
    x_k=1-\sum_{i\neq k}x_i,
\end{equation}
\noindent and
\begin{equation}\label{bond_length_dep}
\begin{split}
\bar{d}_{111} =\ & \sum_{i\neq k}\sum_{j\neq k}x_ix_jd^{ij}_{111} + \sum_{i\neq k} x_i\biggl(1-\sum_{j\neq k}x_j\biggr)d^{ik}_{111} \\
& + \sum_{i\neq k}\biggl(1-\sum_{j\neq k}x_j\biggr)x_id^{ki}_{111} + \biggl(1-\sum_{j\neq k}x_j\biggr)^{\!2}d^{kk}_{111}.
\end{split}
\end{equation}
\noindent Differentiating \Cref{bond_length_dep} with respect to independent composition variable $x_1$ gives
\begin{small}
\begin{equation}\label{bond_length_der_dep}
\begin{split}
&\frac{\partial \bar{d}_{111}}{\partial x_1}=\sum_{i\neq k}(x_id^{i1}_{111}+x_id^{1i}_{111})-\sum_{i\neq 1,k}(x_id^{ik}_{111}+x_id^{ki}_{111})\\ & +(1-2x_1-\sum_{i\neq 1,k}x_i)d^{1k}_{111}+ (1-2x_1-\sum_{i\neq 1,k}x_i)d^{k1}_{111} - 2(1-\sum_{i\neq k} x_i)d^{kk}_{111}\\&
=2\left(\sum_{i\neq k}x_id^{i1}_{111} - \sum_{i\neq 1,k} x_id^{ik}_{111} + (x_k-x_1)d^{1k}_{111}-x_kd^{kk}_{111}\right).
\end{split}
\end{equation}
\end{small}
\noindent Differentiating \Cref{lattice_param} and plugging in the result from \Cref{bond_length_der_dep}, we have
\begin{equation}
\frac{\partial \bar{a}}{\partial x_1}=\frac{4}{\sqrt{3}}\left(\sum_{i\neq k}x_id^{i1}_{111} - \sum_{i\neq 1,k} x_id^{ik}_{111} + (x_k-x_1)d^{1k}_{111}-x_kd^{kk}_{111}\right).
\end{equation}
Additionally, the cubic elastic constants $C_{11}$, $C_{12}$, and $C_{44}$ were computed for each composition of interest using the \texttt{CubicElasticConstantsAnalyzer} within the MaterialsFramework~\cite{sariturk_2025_15731044}. $4\times4\times4$ BCC SQS supercells were first relaxed using the ORB v2 universal MLIP~\cite{neumann2024orb} with unconstrained cell shape to a force convergence of $f_{\mathrm{max}} < \SI{0.01}{\electronvolt/\angstrom}$. Three finite strain modes were then applied to the relaxed cells: a uniform volumetric distortion to extract the bulk modulus $B$, an orthorhombic distortion to isolate $C'=\left(C_{11}-C_{12}\right)/2$, and a monoclinic shear to determine $C_{44}$. Polycrystalline-averaged elastic properties were then obtained via the Voigt-Reuss-Hill (VRH) scheme, yielding the macroscopically isotropic Young's modulus $E$ and Poisson's ratio $\nu$.

\subsection{Verifying Microstructure Instability with Phase Field Modeling}
\label{sec:pf_theory}

While the eigenanalysis of the Gibbs energy Hessian described in \Cref{sec:stability_theory} identifies whether a composition is spinodally unstable and in which compositional direction decomposition is favored, it does not reveal the temporal and morphological evolution of the resulting microstructure. To bridge this gap, we couple the CALPHAD-derived thermodynamics directly to a phase field simulation based on the Cahn-Hilliard equation~\cite{cahn1961spinodal,cahn1958free}.

\subsubsection{Free Energy Functional and Governing Equations}

For an $n$-component alloy described by $n-1$ independent composition fields $x_i(\boldsymbol{r},t)$, $i=1,\ldots,n-1$, with the $n$-th component serving as the reference (dependent variable), the total free energy functional is
\begin{equation}\label{eq:pf_functional}
    \mathcal{F}[\boldsymbol{x}]
    = \int_\Omega \left[
        G^\alpha_m(\boldsymbol{x})
        + \sum_{i=1}^{n-1} \frac{\kappa_i}{2}\,\bigl|\nabla x_i\bigr|^2
      \right] dV,
\end{equation}
where $G^\alpha_m(\boldsymbol{x})$ is the molar Gibbs energy of phase $\alpha$ (the CALPHAD-derived quantity described in \Cref{sec:gibbs_models}), and $\kappa_i$ (\si{\joule\metre\squared\per\mol}) is the gradient energy coefficient for field $i$ that penalizes sharp compositional interfaces.  The time evolution of each composition field is governed by the Cahn-Hilliard equation:
\begin{equation}\label{eq:CH_multicomp}
    \frac{\partial x_i}{\partial t}
    = \nabla \cdot \sum_{j=1}^{n-1} M_{ij}(\boldsymbol{x})\,
      \nabla \mu_j^{\rm tot},
    \qquad i = 1,\ldots,n-1,
\end{equation}
where $\boldsymbol{M}(\boldsymbol{x})$ is the composition-dependent mobility tensor (\si{\metre\tothe{5}\per\joule\per\second}) and $\mu_j^{\rm tot}$ is the variational chemical potential:
\begin{equation}\label{eq:mu_var_pf}
    \mu_j^{\rm tot}
    = \frac{\delta\mathcal{F}}{\delta x_j}
    = \underbrace{\frac{\partial G^\alpha_m}{\partial x_j}}_{\textstyle\mu_j - \mu_n}
      \;-\; \kappa_j\,\nabla^2 x_j.
\end{equation}
The bulk driving force $\mu_j - \mu_n = \partial G^\alpha_m/\partial x_j$ is precisely the relative chemical potential already established in \Cref{eq:first_derivative}.  To prevent diffusion from diverging as any component approaches a pure-element limit, we use a degenerate mobility,
\begin{equation}\label{eq:mobility_pf}
    M_{ij}(\boldsymbol{x}) = \mathring{M}_{ij}\prod_{k=1}^{n} x_k,
\end{equation}
where $\mathring{M}_{ij}$ is the composition-independent (bare) mobility matrix. The product over all $n$ mole fractions vanishes whenever any $x_k \to 0$, enforcing physical boundary behavior.

\Cref{eq:CH_multicomp} is solved numerically on a uniform $N_g \times N_g$ grid (domain size $L_x$) using a pseudo-spectral semi-implicit scheme \cite{chen1998applications}. Diagonal mobility terms are treated implicitly (providing unconditional damping of high-wavenumber modes) while off-diagonal cross-terms are treated explicitly:
\begin{equation}\label{eq:semi_implicit}
    \hat{x}_i^{\,n+1}
    = \frac{
        \hat{x}_i^{\,n}
        + \Delta t\,k^2 \displaystyle\sum_j M^{\rm eff}_{ij}\,\hat{\mu}_{j,\rm bulk}^n
        - \Delta t\,k^4 \displaystyle\sum_{j \neq i} M^{\rm eff}_{ij}\,\kappa_j\,\hat{x}_j^{\,n}
      }{
        1 + \Delta t\,k^4\,M^{\rm eff}_{ii}\,\kappa_i
      },
\end{equation}
where $\hat{(\cdot)}$ denotes the discrete Fourier transform, $k^2$ is the squared wavenumber, and $M^{\rm eff}_{ij} = \mathring{M}_{ij}\langle\prod_k x_k\rangle$ is the spatially-averaged effective mobility.  The stable time step is estimated from the Hessian of the free energy at the mean composition $\boldsymbol{x}_0$:
\begin{equation}\label{eq:dt_stable_pf}
    \Delta t_{\rm stable}
    = \frac{\alpha_{\rm s}\,\Delta x^2}
           {\max_i(\mathring{M}_{ii})\,|\lambda_{\rm max}(H)|},
\end{equation}
where $\alpha_{\rm s} < 1$ is a safety factor, $\Delta x = L_x/N_g$, and $|\lambda_{\rm max}(H)|$ is the largest absolute eigenvalue of the Gibbs energy Hessian $H_{ij} = \partial^2 G^\alpha_m/\partial x_i\partial x_j$ evaluated in a neighborhood of $\boldsymbol{x}_0$.

\subsubsection{Eyre Convex-Splitting Stabilization}
\label{sec:eyre}

The semi-implicit scheme of \Cref{eq:semi_implicit} treats the gradient energy term $\kappa_i k^4$ implicitly, which unconditionally damps high-wavenumber oscillations arising from interfacial stiffness. However, within the spinodal region the diagonal Hessian elements $H_{ii}$ are \emph{negative}, and the corresponding bulk contribution $M_{ii} H_{ii} k^2$ enters the numerator with a destabilizing sign. For large $|\Delta t H_{ii}|$ this explicit term can cause unbounded growth and numerical blowup, regardless of the spatial resolution.

Following Eyre \cite{eyre1998unconditionally}, we apply a convex-splitting decomposition of the molar Gibbs energy $G^\alpha_m$ into a \emph{contracting} (convex) part $G^+$ and an \emph{expanding} (concave) part $G^-$,
\begin{equation}\label{eq:eyre_split}
    G^\alpha_m(\boldsymbol{x}) = G^+(\boldsymbol{x}) - G^-(\boldsymbol{x}),
    \qquad G^{\pm} \text{ convex},
\end{equation}
and treat $G^+$ implicitly and $G^-$ explicitly in time. For a quadratic approximation to the bulk free energy at the mean composition $\boldsymbol{x}_0$, a sufficient convex-splitting for component $i$ is
\begin{equation}\label{eq:eyre_Adef}
    G^+_i = \tfrac{A_i}{2}\,x_i^2,
    \qquad
    G^-_i = \tfrac{A_i}{2}\,x_i^2 - G^\alpha_m(\boldsymbol{x}),
\end{equation}
where the \emph{stabilization constant} $A_i$ must satisfy
\begin{equation}\label{eq:eyre_condition}
    A_i \;\geq\; \max \bigl(0,\,-H_{ii}(\boldsymbol{x}_0)\bigr)
\end{equation}
to ensure that $G^+_i - \tfrac{A_i}{2}x_i^2$ remains convex \cite{eyre1998unconditionally,shen2010numerical}.  With this splitting, the update rule for field $i$ in Fourier space becomes
\begin{equation}\label{eq:eyre_update}
\resizebox{\linewidth}{!}{$\displaystyle
    \hat{x}_i^{\,n+1}
    = \frac{
        \hat{x}_i^{\,n}
        + \Delta t\,k^2 \sum_j M^{\rm eff}_{ij}\,\hat{\mu}_{j,\rm bulk}^{\,n}
        - \Delta t\,k^2\,M^{\rm eff}_{ii}\,A_i\,\hat{x}_i^{\,n}
        - \Delta t\,k^4 \sum_{j \neq i} M^{\rm eff}_{ij}\,\kappa_j\,\hat{x}_j^{\,n}
      }{
        1 + \Delta t\,k^2\,M^{\rm eff}_{ii}\,A_i
        + \Delta t\,k^4\,M^{\rm eff}_{ii}\,\kappa_i
      }
$},
\end{equation}
where the diagonal chemical potential term $M^{\rm eff}_{ii}\,\hat{\mu}_{i,\rm bulk}^{\,n}$ and off-diagonal bulk cross-terms $\sum_{j\neq i}M^{\rm eff}_{ij}\,\hat{\mu}_{j,\rm bulk}^{\,n}$ both enter at order $k^2$, while the off-diagonal gradient cross-terms enter at order $k^4$, consistent with \Cref{eq:semi_implicit}. The denominator is strictly positive for any $\Delta t > 0$, $A_i \geq 0$, $\kappa_i > 0$, making the scheme \emph{unconditionally stable}: the solution cannot grow without bound regardless of the chosen time step.

\subsubsection{CALPHAD-Coupled Surrogate for Thermodynamic Driving Forces}
\label{sec:calphad_surrogate}

Direct evaluation of PyCalphad at every spatial grid point and every time step is computationally prohibitive as each equilibrium call requires an iterative minimization. Instead, all thermodynamic quantities are pre-computed on a dense composition grid spanning the relevant simplex, stored in an HDF5 file, and replaced by smooth surrogate functions that are evaluated in $\mathcal{O}(\log N)$ time during simulation.

At each grid point $\boldsymbol{x}^{(k)}$, three quantities are evaluated once using PyCalphad's \texttt{IsolatedPhase} framework, which constrains every calculation to the single-phase BCC free energy surface: the molar Gibbs energy $G^{\alpha,(k)}_m$ (\si{\joule\per\mol}); the absolute chemical potential $\mu^{(k)}_i$ of each element $i$ (\si{\joule\per\mol}); and the Jansson second derivative $\Lambda^{(k)}_{ij} = \partial\mu_i/\partial x_j$ \cite{kunselman2024analytically} (\si{\joule\per\mol}), which is the analytic derivative of the chemical potential with respect to composition at single-phase equilibrium. All three datasets are stored in a structured HDF5 file and loaded once at the start of a simulation run; no further thermodynamic calculations are performed during time integration.

The surrogate is constructed using composition-appropriate interpolation backends: a cubic spline for binary systems ($n=2$), a $C^1$-smooth Clough-Tocher triangular interpolant \cite{clough1965finite} for ternary systems ($n=3$), and a two-stage thin-plate-spline RBF-to-regular-grid scheme \cite{virtanen2020scipy} for quaternary and higher-order systems ($n \geq 4$). All three methods share the same surrogate class and are used identically for $G^\alpha_m$, $\mu_i$, and $\Lambda_{ij}$.

\subsubsection{Derivative Methods for the Bulk Chemical Potential}
\label{sec:deriv_methods_pf}

Computing the bulk driving force $\mu_j - \mu_n = \partial G^\alpha_m/\partial x_j$ and the Hessian $H_{ij}$ from the surrogate can be done in three ways, selectable in the interface. The choice matters because the Hessian controls both the spinodal wavelength $\lambda^* = 4\pi\sqrt{\kappa_i/|H_{ij}|}$ and the stable time step (\Cref{eq:dt_stable_pf}), so derivative errors propagate directly into microstructural length and time scales.

\paragraph{Method I — Finite differences on $G^\alpha_m$}
The chemical potential and Hessian are obtained numerically from the $G^\alpha_m$ surrogate by central finite differences:
\begin{equation}\label{eq:fd_mu_pf}
    \left(\mu_j - \mu_n\right)^{\rm fd}
    = \frac{G^\alpha_m(\boldsymbol{x} + \varepsilon\boldsymbol{e}_j)
            - G^\alpha_m(\boldsymbol{x} - \varepsilon\boldsymbol{e}_j)}
           {2\varepsilon}
    + O(\varepsilon^2),
\end{equation}
and similarly for $H_{ij}$ using the standard second-order stencil, where $\boldsymbol{e}_j$ is the unit vector in the $j$-th composition direction and $\varepsilon = 10^{-4}$. This approach requires only the $G^\alpha_m$ table. No chemical potential data need be stored. However, truncation error compounds with the smoothing bias of the surrogate, and errors in $H_{ij}$ are amplified near the simplex boundaries where the ideal-entropy contribution $-RT\sum_i x_i \ln x_i$ makes $G^\alpha_m$ highly curved.

\paragraph{Method II — Direct Jansson interpolation}
Under the simplex constraint $x_n = 1 - \sum_{i=1}^{n-1} x_i$, the Gibbs-Duhem equation gives \cite{de1979configurational}:
\begin{equation}\label{eq:mu_jansson_pf}
    \mu_j - \mu_n
    = \frac{\partial G^\alpha_m}{\partial x_j}\bigg|_{\rm simplex},
\end{equation}
and the Hessian element is (from \Cref{eq:second_derivatives}):
\begin{equation}\label{eq:H_jansson_pf}
    H_{ij}
    = \frac{\partial^2 G^\alpha_m}{\partial x_i\,\partial x_j}\bigg|_{\rm simplex}
    = \Lambda_{ij} - \Lambda_{nj},
\end{equation}
where $\Lambda_{ij} = \partial\mu_i/\partial x_j$ is the Jansson second derivative \cite{kunselman2024analytically} pre-computed and stored in the HDF5 table. Separate surrogate functions are constructed for each $\mu_j - \mu_n$ (from \Cref{eq:mu_jansson_pf}) and each $H_{ij}$ (from \Cref{eq:H_jansson_pf}) using identical backends. This method is free of finite-difference truncation error and satisfies the Gibbs-Duhem relation exactly at every queried composition. For an $n$-component system, it requires $(n-1)$ additional surrogates for the chemical potentials and $(n-1)^2$ for the Hessian components, a total of $(n-1)n$ interpolants beyond the $G^\alpha_m$ surface, all constructed once at simulation startup.

\paragraph{Method III — Hybrid: Jansson $\mu$, finite-difference $H$}
The bulk driving force $\mu_j - \mu_n$ is obtained from the Jansson surrogates (\Cref{eq:mu_jansson_pf}), while the Hessian $H_{ij}$ is obtained by finite differences on $G^\alpha_m$ (\Cref{eq:fd_mu_pf}). This combination is appropriate when the Jansson derivative dataset (\texttt{dMU}) is unavailable or when PyCalphad's numerical Hessian data are unreliable near a phase boundary, while still providing thermodynamically consistent chemical potentials for the time-evolution equation.

\subsubsection{Validation of Derivative Methods}

\begin{figure*}[!tp]
    \centering
    \includegraphics[width=\textwidth]{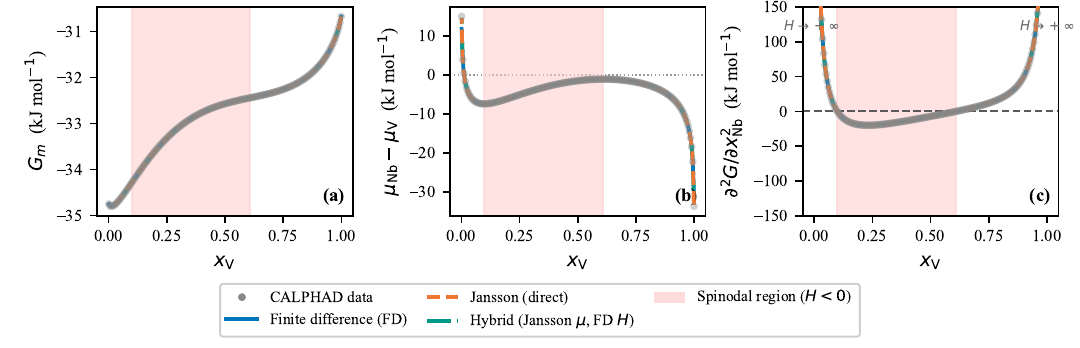}
    \caption{Comparison of the three derivative methods for the \ch{Nb-V} BCC system at $T = \SI{500}{\celsius}$. Grey markers: raw CALPHAD data. (a) Molar Gibbs energy $G^\alpha_m$: all three modes are identical (shared surrogate). (b) Relative chemical potential $\mu_{\rm Nb} - \mu_{\rm V}$: \texttt{fd} deviates by $\mathrm{RMSE} = \SI{302}{\joule\per\mol}$; \texttt{jansson} and \texttt{hybrid} are exact. (c) Hessian $\partial^2 G^\alpha_m/\partial x_{\rm Nb}^2$ (clipped to $\pm\SI{150}{\kilo\joule\per\mol}$; divergence at the pure-element limits is annotated): \texttt{fd} RMSE $= \SI{442}{\kilo\joule\per\mol}$; \texttt{jansson} is exact. The shaded pink band marks the spinodal region ($H < 0$) identified from the \texttt{jansson} Hessian.}
    \label{fig:deriv_methods}
\end{figure*}

\Cref{fig:deriv_methods} compares the three modes against raw CALPHAD data for the \ch{Nb-V} BCC binary system at $T = \SI{500}{\celsius}$ as a function of the vanadium mole fraction $x_{\rm V}$. Panel (a) confirms that the molar Gibbs energy $G^\alpha_m$ is identical across all modes, since it is always evaluated from the same $G^\alpha_m$ surrogate. Panel~(b) shows the relative chemical potential $\mu_{\rm Nb} - \mu_{\rm V}$: the \texttt{fd} mode deviates visibly from the CALPHAD reference (RMSE $= \SI{302}{\joule\per\mol}$) while \texttt{jansson} and \texttt{hybrid} reproduce the stored values exactly (RMSE $= \SI{0}{\joule\per\mol}$). Panel~(c) shows the most striking difference: the Hessian $\partial^2 G^\alpha_m/\partial x_{\rm Nb}^2$ is displayed with the $y$-axis clipped to $\pm\SI{150}{\kilo\joule\per\mol}$ (the ideal-entropy divergence at $x_{\rm V} \to 0$ or $1$ is annotated); the \texttt{fd} mode yields an RMSE of $\SI{442}{\kilo\joule\per\mol}$ relative to the CALPHAD reference, entirely due to truncation error amplification near the simplex boundaries, whereas \texttt{jansson} matches exactly. Since the spinodal decomposition wavelength scales as $\lambda^* \propto |H_{ij}|^{-1/2}$, Hessian errors of this magnitude would translate into erroneous microstructural length-scale predictions. The \texttt{jansson} method eliminates this source of error.

\begin{table*}[!tp]
    \centering
    \footnotesize
    \caption{Summary of derivative methods for bulk chemical potentials and Hessian in the CALPHAD-coupled Cahn-Hilliard model. RMSE values are for the \ch{Nb-V} BCC binary at $T = \SI{500}{\celsius}$ ($n = 200$ composition grid points) showing exact agreement with CALPHAD data.}
    \label{tab:deriv_methods}
    \setlength{\tabcolsep}{7pt}
    \begin{tabularx}{0.90\textwidth}{lllXcc}
        \toprule
        Mode & $\mu_j - \mu_n$ source & $H_{ij}$ source
             & HDF5 data required
             & RMSE$_{\mu}$ (\si{\joule\per\mol})
             & RMSE$_{H}$ (\si{\joule\per\mol}) \\
        \midrule
        \texttt{fd}
            & FD on $G^\alpha_m$ (Eq.~\ref{eq:fd_mu_pf})
            & FD on $G^\alpha_m$
            & \texttt{/GM}
            & 302
            & 442\,219 \\
        \texttt{jansson}
            & Eq.~\ref{eq:mu_jansson_pf}
            & Eq.~\ref{eq:H_jansson_pf}
            & \texttt{/GM}, \texttt{/MU}, \texttt{/dMU}
            & $\boldsymbol{0}$
            & $\boldsymbol{0}$ \\
        \texttt{hybrid}
            & Eq.~\ref{eq:mu_jansson_pf}
            & FD on $G^\alpha_m$
            & \texttt{/GM}, \texttt{/MU}
            & $\boldsymbol{0}$
            & 442\,219 \\
        \bottomrule
    \end{tabularx}
\end{table*}

\begin{figure*}[!tp]
    \centering
    \includegraphics[width=\textwidth]{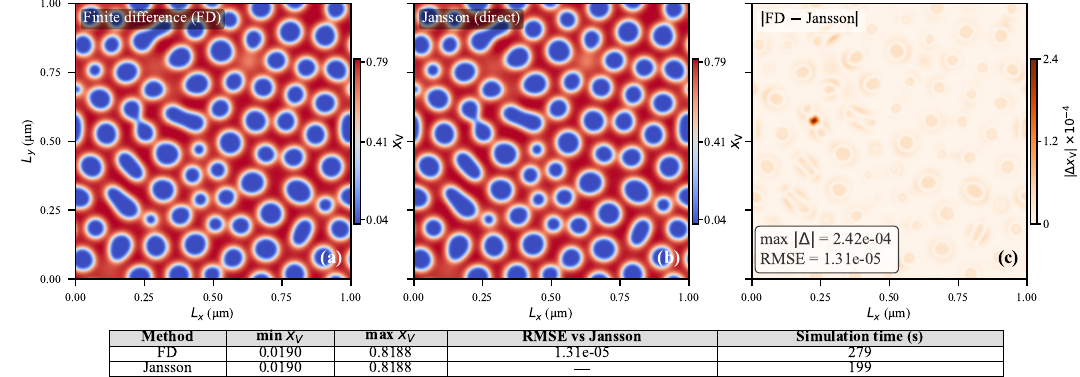}\\[6pt]
    \includegraphics[width=\textwidth]{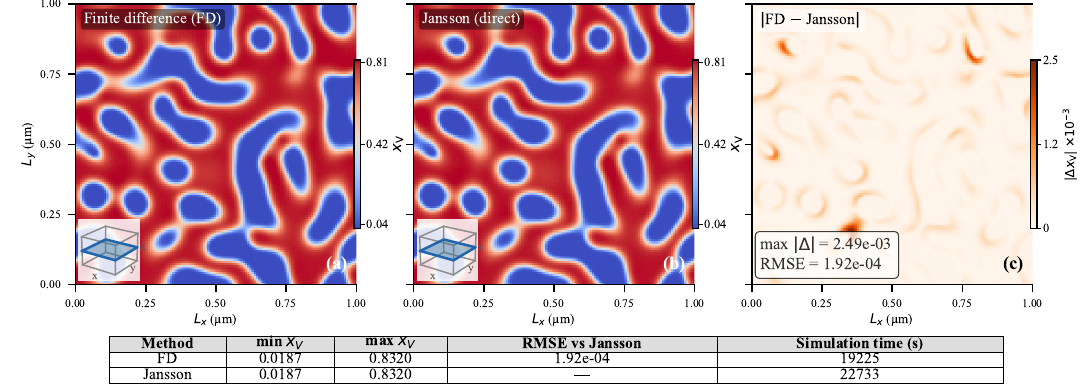}
    \caption{Derivative-method comparison (\texttt{fd} vs.\ \texttt{jansson}) for \ch{Nb-V} BCC\textsubscript{A2} at $T = \SI{500}{\celsius}$ ($c_0 = x_{\rm V} = 0.50$, $L = \SI{1}{\micro\metre}$). \textit{Top row} (2D, $256\times256$ grid, $t = \SI{800}{\second}$, 80\,000 steps): (a)~finite-difference and (b)~Jansson direct composition fields $x_{\rm V}$; (c)~absolute difference map $|x^{\rm fd}_{\rm V} - x^{\rm jansson}_{\rm V}|$ ($\mathrm{RMSE} = \num{1.31e-5}$, max $= \num{2.42e-4}$); table: composition statistics and wall times (FD: \SI{279}{\second}; Jansson: \SI{199}{\second}). \textit{Bottom row} (3D, $64^3$ grid, $t = \SI{800}{\second}$, 5\,000 steps, $z = L/2$ mid-plane slice): (a)~finite-difference, (b)~Jansson, (c)~absolute difference; discrepancies remain confined to domain interfaces in both 2D and 3D, confirming method equivalence independent of dimensionality.}
    \label{fig:microstructure_methods}
    \label{fig:microstructure_methods_3d}
\end{figure*}

To confirm that point-wise derivative agreement translates into equivalent field evolution, we ran two identical Cahn-Hilliard simulations --- one each for the \texttt{fd} and \texttt{jansson} modes --- on the \ch{Nb-V} BCC binary at $T = \SI{500}{\celsius}$ using a $256 \times 256$ periodic domain of side $L = \SI{1}{\micro\metre}$, $\kappa = \SI{1e-12}{\joule\per\metre}$, $M = \SI{5e-21}{\metre\cubed\second\per\kg}$, and $80\,000$ time steps of $\Delta t = \SI{0.01}{\second}$ ($t = \SI{800}{\second} \approx 26\,\tau$, where $\tau \approx \SI{31}{\second}$ is the spinodal growth time constant at the dominant wavenumber $k^* = \sqrt{|H_{\min}|/4\kappa}$). Both runs were seeded identically. The only variable was the derivative method used to evaluate the chemical-potential driving force $\delta F/\delta c$. \Cref{fig:microstructure_methods} presents the final $x_{\rm V}$ composition fields and the absolute per-pixel difference map. After $\SI{800}{\second}$ of evolution the field is fully phase-separated, with well-coarsened Nb-rich ($x_{\rm V} \approx 0.02$) and V-rich ($x_{\rm V} \approx 0.82$) domains of characteristic size $\sim \SI{200}{\nano\metre}$. The two micrographs are visually identical. \texttt{fd} shows a small residual relative to \texttt{jansson} (RMSE $= \num{1.3e-5}$, max $|\Delta x_{\rm V}| = \num{2.4e-4}$) concentrated at domain interfaces where the composition gradient is steepest and FD truncation error is most amplified. This tiny residual does not affect the phase morphology or domain statistics, confirming that both derivative modes are effectively equivalent for the binary system. Differences may grow for higher-order systems where scattered-data interpolants are used. Those cases are addressed in \Cref{sec:quaternary}.

\subsubsection{Elastic Coupling via Coherency Strain}
\label{sec:elastic_coupling_pf}

Coherency strains arising from composition-dependent lattice mismatch contribute an elastic free energy that supplements the chemical driving force. Following Gururajan and Abinandanan \cite{gururajan2007phase}, the total chemical potential entering \Cref{eq:CH_multicomp} is augmented by an elastic term:
\begin{equation}\label{eq:mu_tot_elastic}
    \mu_j^{\rm tot} = \mu_j^{\rm ch} + \mu_j^{\rm el},
\end{equation}
where $\mu_j^{\rm ch} = \partial G^\alpha_m/\partial x_j - \kappa_j\nabla^2 x_j$ is the chemical contribution (\Cref{eq:mu_var_pf}) and $\mu_j^{\rm el}$ is the elastic contribution derived from the variation of the elastic strain energy with respect to composition.

Mechanical equilibrium is solved at each elastic update step using the Lippmann-Schwinger spectral scheme of Moulinec and Suquet \cite{moulinec1998numerical}, which reformulates the Navier equation as a Fredholm integral equation and solves it iteratively in Fourier space. A homogeneous reference stiffness $\mathbf{C}^0$ is chosen to accelerate convergence, and the spatially varying stiffness field $\mathbf{C}(\boldsymbol{r})$ is treated as a perturbation. The elastic state of the microstructure is determined by the eigenstrain field $\boldsymbol{\varepsilon}^*(\boldsymbol{r})$, which encodes the stress-free lattice mismatch between the local composition and a reference state. Two formulations are implemented in the present framework, summarized in \Cref{tab:eigenstrain_models}.

\begin{table*}[!tp]
    \centering
    \caption{Comparison of the two eigenstrain models available in the CALPHAD-coupled phase field solver. $\alpha_k$ denotes the Vegard coefficient for element $k$, $a(\boldsymbol{x})$ the SQS lattice parameter at composition $\boldsymbol{x}$, $a_0 = a(\boldsymbol{x}_0)$ the reference lattice parameter at the parent composition, $\boldsymbol{\sigma}$ the Cauchy stress, and $N_V$ the molar density. All models use the Moulinec-Suquet spectral scheme with homogeneous reference stiffness $\mathbf{C}^0$; the stiffness contrast correction is neglected in both cases, which is standard practice when the contrast is moderate.}
    \label{tab:eigenstrain_models}
    \renewcommand{\arraystretch}{1.35}
    \small
    \begin{tabular}{lp{4.2cm}p{3.6cm}p{3.4cm}}
        \toprule
        & \textbf{Classical} (\texttt{fixed\_vegard})
        & \multicolumn{2}{l}{\textbf{Composition-dependent} (\texttt{local\_vegard})} \\
        \midrule
        Eigenstrain field
        & $\boldsymbol{\varepsilon}^*(\boldsymbol{r}) = \displaystyle\sum_k \alpha_k\, x_k(\boldsymbol{r})\,\mathbf{I}$
        & \multicolumn{2}{l}{$\boldsymbol{\varepsilon}^*(\boldsymbol{r}) = \dfrac{a(\boldsymbol{x}(\boldsymbol{r})) - a_0}{a_0}\,\mathbf{I}$} \\[6pt]
        Elastic $\mu_k^{\rm el}$
        & $-\boldsymbol{\sigma}(\boldsymbol{r}):\boldsymbol{\Lambda}_k\,/\,N_V$
        & \multicolumn{2}{l}{$-\alpha_k(\boldsymbol{x}_0)\,\mathrm{tr}(\boldsymbol{\sigma}(\boldsymbol{r}))\,/\,N_V$} \\[4pt]
        Reference stiffness $\mathbf{C}^0$
        & $\tfrac{1}{2}(\mathbf{C}_{\rm mat} + \mathbf{C}_{\rm ppt})$
        & \multicolumn{2}{l}{$\mathbf{C}(\boldsymbol{x}_0)$ from SQS interpolant} \\[4pt]
        Stiffness field $\mathbf{C}(\boldsymbol{r})$
        & $(1-h)\,\mathbf{C}_{\rm mat} + h\,\mathbf{C}_{\rm ppt}$
        & \multicolumn{2}{l}{SQS interpolant evaluated pointwise at $\boldsymbol{x}(\boldsymbol{r})$} \\[4pt]
        $\alpha_k$ evaluation
        & Fixed at tie-line midpoint (step 1b)
        & \multicolumn{2}{l}{$\partial\ln a/\partial x_k$ at $\boldsymbol{x}_0$ via finite difference on SQS interpolant} \\[4pt]
        Initial $\boldsymbol{\varepsilon}^*$ (uniform $\boldsymbol{x}_0$)
        & $\sum_k \alpha_k\, x_{k,0}\,\mathbf{I} \neq \mathbf{0}$
        & \multicolumn{2}{l}{$\mathbf{0}$ (vanishes exactly)} \\
        \bottomrule
    \end{tabular}
\end{table*}

In the classical model (\texttt{fixed\_vegard}), the eigenstrain is linear in the composition fields with fixed Vegard coefficients $\alpha_k = \partial\ln a/\partial x_k$ evaluated at the midpoint between the tie-line endpoints. The elastic chemical potential follows directly from the variation of the elastic energy with respect to $x_k$: $\mu_k^{\rm el} = -\boldsymbol{\sigma}:\boldsymbol{\Lambda}_k/N_V$, where $\boldsymbol{\Lambda}_k = \alpha_k\mathbf{I}$ is the Vegard tensor for element $k$. The reference stiffness is taken as the arithmetic mean of the endpoint stiffness tensors, $\mathbf{C}^0 = \tfrac{1}{2}(\mathbf{C}_{\rm mat}+\mathbf{C}_{\rm ppt})$, and the spatially varying stiffness is interpolated between the two endpoints using a smooth switching function $h(x_{\rm ref})$.

The composition-dependent model (\texttt{local\_vegard}) removes the assumption of fixed $\alpha_k$ by computing the eigenstrain directly from the MLIP-derived SQS lattice parameter $a(\boldsymbol{x})$ interpolated across the full composition space. The eigenstrain at each point is the volumetric mismatch relative to the parent composition: $\varepsilon^*(\boldsymbol{r}) = [a(\boldsymbol{x}(\boldsymbol{r})) - a_0]/a_0$, which vanishes identically in the initial uniform microstructure and grows only as phase separation develops. This is physically correct: an undecomposed single-phase alloy carries no coherency stress. The classical model, by contrast, assigns a nonzero eigenstrain proportional to $\boldsymbol{x}_0$ everywhere in the initial field. The reference stiffness $\mathbf{C}^0 = \mathbf{C}(\boldsymbol{x}_0)$ is evaluated from the SQS interpolant at the parent composition, capturing the correct elastic environment of the parent phase, while the spatially varying stiffness $\mathbf{C}(\boldsymbol{r})$ is evaluated pointwise from the same interpolant. The elastic chemical potential uses $\alpha_k(\boldsymbol{x}_0)$ --- the Vegard coefficient at the parent composition computed by finite difference on the SQS interpolant --- rather than a midpoint value, which is a first-order approximation that avoids per-point gradient evaluations on the scattered interpolant.

In both models, the stiffness-contrast correction ($\tfrac{1}{2}\delta\mathbf{C}:\boldsymbol{\varepsilon}^{\rm el}:\boldsymbol{\varepsilon}^{\rm el}$) is neglected, as is standard in spectral phase field codes when the elastic contrast between the two phases is moderate \cite{gururajan2007phase}. The mechanical equilibrium is re-solved every $n_{\rm el}$ time steps and the resulting elastic chemical potential is cached between solves to reduce computational cost.

\section{Investigating Spinodal Decomposition in the \ch{Hf-Nb-Ti-V} System}

\subsection{Microstructure Stability Analysis}
In this section, we use our stability map framework to investigate spinodal decomposition in the BCC phase of the \ch{Hf-Nb-Ti-V} system. In a recent experimental study, single-phase BCC was observed for \ch{Hf_{10}Nb_{40}Ti_{40}V_{10}}, and spinodal decomposition was reported for an alloy at the equiatomic composition \cite{an2021spinodal}. High lattice strains around \SI{5}{\percent} were also reported for one of the phases resulting from spinodal decomposition, indicating that accounting for coherent strain is important when modeling spinodal decomposition in this system. 

\begin{figure*}[!h]
\centering
\includegraphics[width=0.8\textwidth]{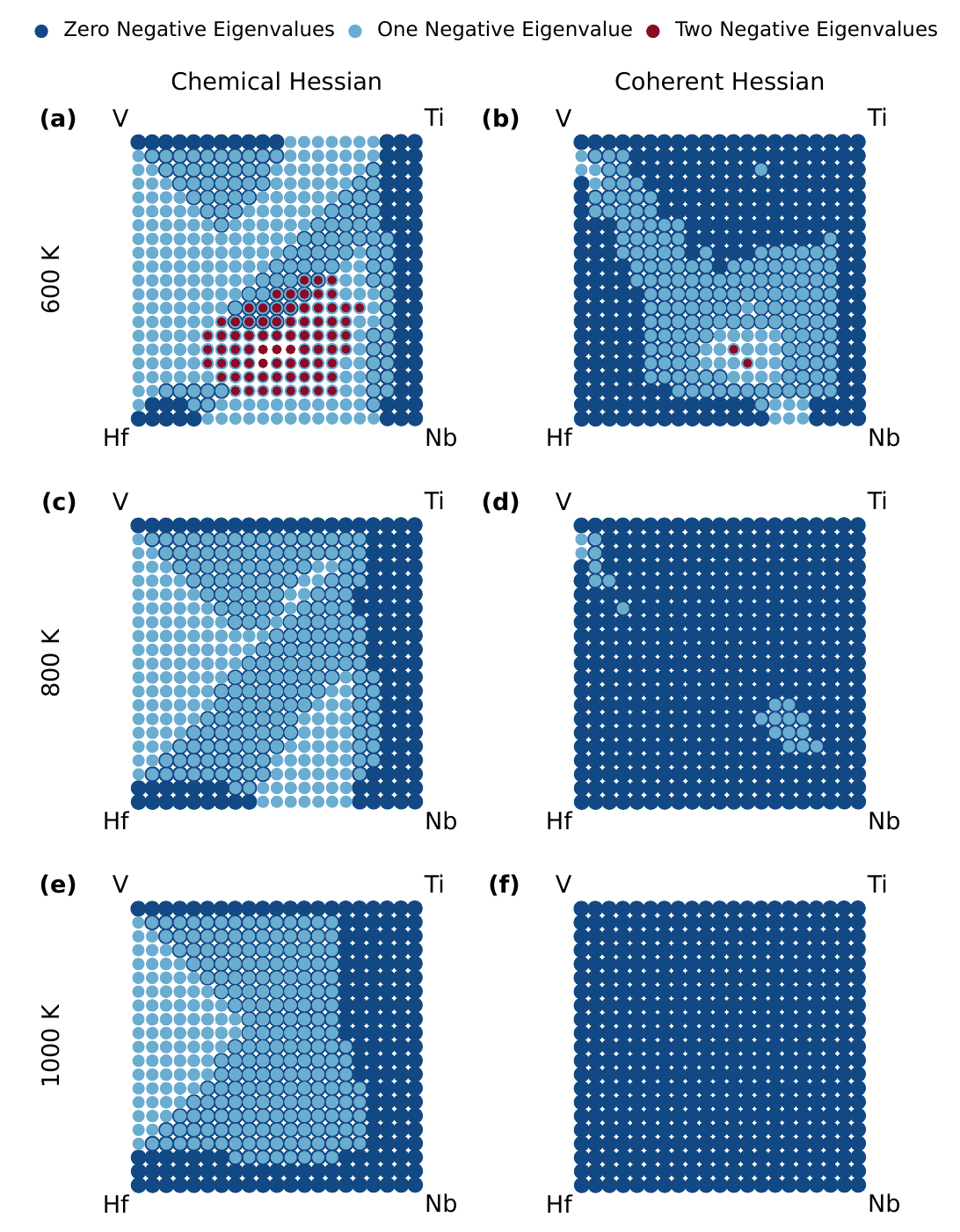}
 \caption{Chemical (a, c, e) and coherent (b, d, f) Hessian stability maps for BCC \ch{Hf-Nb-Ti-V} at \SI{600}{\kelvin}, \SI{800}{\kelvin}, and \SI{1000}{\kelvin}. Composition is projected from the quaternary simplex onto a 2D square with corners Hf, Nb, Ti, and V. The projection is not one-to-one: more than one composition can map to the same point. Points are colored by the number of negative eigenvalues of the corresponding Hessian: zero (dark blue, stable), one (light blue), or two (dark red).}
\label{fig:stability maps}
\end{figure*}

Through the procedure outlined in the previous section, we calculated eigenvalues for the chemical and coherent Gibbs energy Hessians at various temperatures with a sampling density in composition space of \SI{5}{\atpercent}. \Cref{fig:stability maps} displays affine projections of these results at \qtylist{600;800;1000}{\kelvin}. Any missing values represent convergence failures, which are most common when the composition of at least one component is dilute. A limitation of affine projections is that multiple compositions can map to the same position. To ensure that marker differences are visible, smaller markers are plotted on top of larger markers (e.g.\ a smaller light blue circle with a darker blue ring around it indicates that compositions mapped to that position exhibit both zero and one negative eigenvalue). At \SI{600}{\kelvin}, spinodal shrinkage due to the addition of coherent strain contributions is most pronounced for compositions close to the \ch{Hf-V} edge at higher Hf content and the \ch{Hf-Ti} edge (which manifests as a diagonal in the affine projection based on the placement of the pure components). The region of two negative eigenvalues from the chemical Hessian almost entirely vanishes when coherency is taken into account. At \SI{800}{\kelvin}, the chemical Hessian predicts spinodal decomposition at almost all compositions except for those rich in Ti and Nb, whereas the coherent Hessian predicts only small regions of instability for compositions close to the \ch{Hf-V} edge at higher V content and the \ch{Nb-V} edge (which is a diagonal). Finally, at \SI{1000}{\kelvin}, the chemical stability map closely resembles that at \SI{800}{\kelvin} with modest shrinkage while all instability is suppressed when coherent contributions are considered.

\begin{figure*}[!h]
\centering
\includegraphics[width=0.9\textwidth]{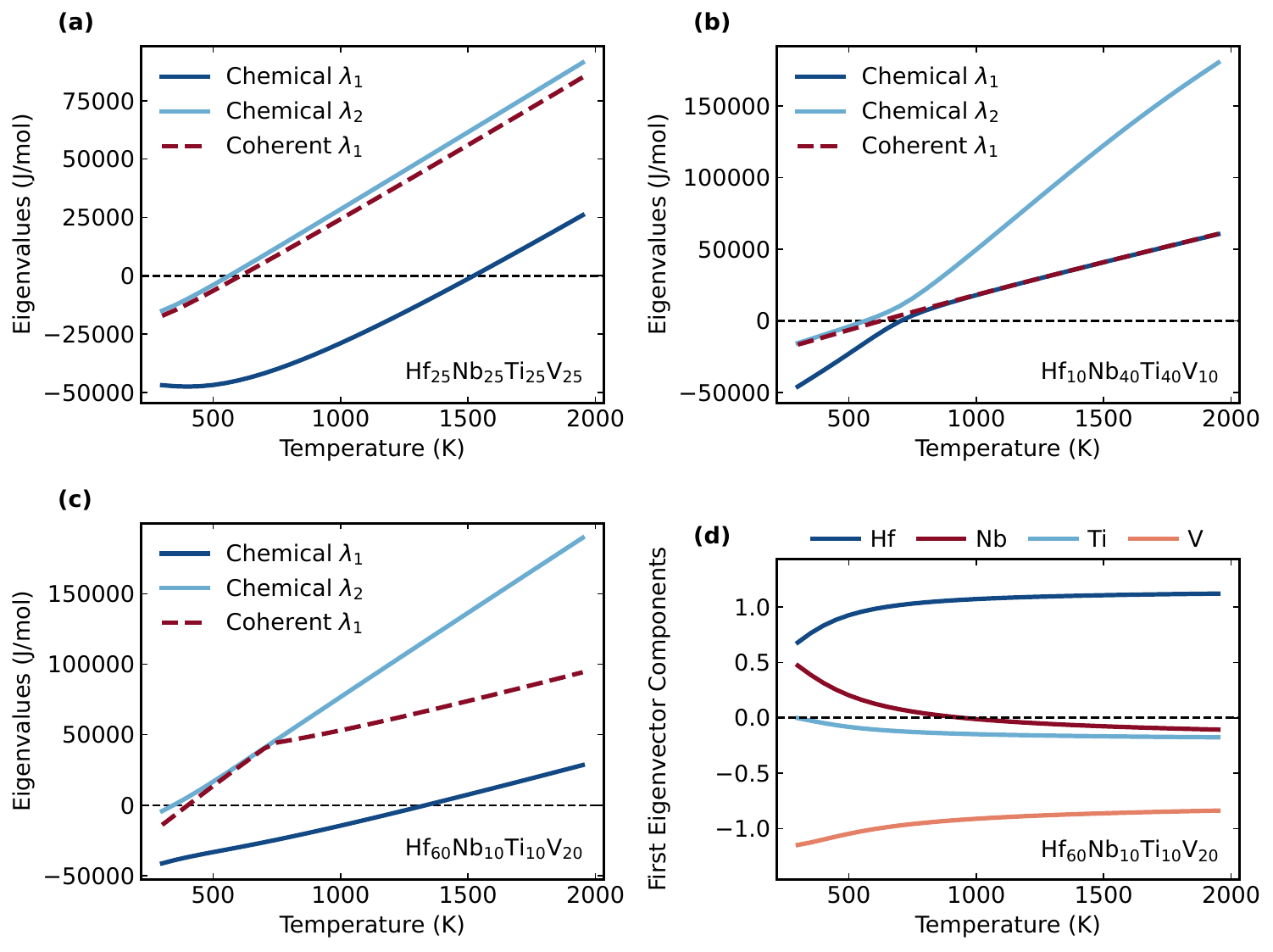}
\caption{Gauge-invariant chemical first and second and coherent first eigenvalues of the BCC Hessian as a function of temperature for (a) \ch{Hf_{25}Nb_{25}Ti_{25}V_{25}}, (b) \ch{Hf_{10}Nb_{40}Ti_{40}V_{10}}, and (c) \ch{Hf_{60}Nb_{10}Ti_{10}V_{20}}. The horizontal dashed line marks the stability threshold, $\lambda_1 = 0$. Panel (d) shows the temperature dependence of the first gauge-invariant eigenvector components of the chemical Hessian for \ch{Hf_{60}Nb_{10}Ti_{10}V_{20}}, decomposed into Hf, Nb, Ti, and V contributions.}
\label{fig:eigenvector}
\end{figure*}

The consistent pattern of coherent strain contributions suppressing the spinodal in regions rich in Hf and low in Ti and Nb could be explained by examining the components of the first eigenvector of the chemical Hessian. \Cref{fig:eigenvector}d displays the components of this vector for a BCC alloy with composition \ch{Hf_{60}Nb_{10}Ti_{10}V_{20}} as a function of temperature from \qtyrange{600}{1000}{\kelvin}. \Cref{fig:eigenvector}c shows that, at this composition, the first eigenvalue of the chemical Hessian is negative for all $T$ below approximately \SI{1500}{\kelvin}, while the coherent Hessian eigenvalues remain positive for all $T$ above approximately \SI{500}{\kelvin}. Thus, in the window $\SI{500}{\kelvin} < T < \SI{1500}{\kelvin}$, the chemical Hessian identifies the eigenvector plotted in \Cref{fig:eigenvector}d as an unstable direction, but coherency stabilizes the alloy against that fluctuation. We see that such an initial composition fluctuation would result in one phase rich in Hf and low in V while the other phase would be rich in V and low in Hf. However, Hf and V have the largest atomic radius mismatch of all four components, and such a composition fluctuation could cause large coherency strains. Consequently, when an elastic contribution describing coherent strain resulting from composition fluctuations is added to the Gibbs energy at this composition, the local curvature of the Gibbs energy is adjusted and the alloy becomes stable to the previously unstable composition fluctuation.

Eigenvalue investigations can also provide more insight. \Cref{fig:eigenvector} plots the first eigenvalue for the chemical and coherent Hessian as a function of temperature at compositions (a) \ch{Hf_{25}Nb_{25}Ti_{25}V_{25}}, (b) \ch{Hf_{10}Nb_{40}Ti_{40}V_{10}}, and (c) \ch{Hf_{60}Nb_{10}Ti_{10}V_{20}}. For the composition rich in Ti and Nb, coherent strain contributions have little effect on the first eigenvalue, and the spinodal temperatures are almost equivalent. As the composition of Hf increases, coherent strain has a greater effect on the spinodal temperature, leading to more stable solid solutions in Hf rich areas of composition space. \Cref{fig:eigenvector}b also includes the second eigenvalue of the chemical Hessian. For this composition, the first (smallest) eigenvalue of the chemical Hessian deviates from the linear behavior that we see in the other plots because of the large difference in the slopes of the directions of principal curvature as a function of temperature. Coherent contributions penalize composition fluctuations corresponding to the direction of the principal curvature with a steeper slope while they have little effect on composition fluctuations in the direction of the principal curvature with a more gentle slope.

\subsection{Binary \ch{Nb-V}: First Demonstration of the MLIP-Informed Pipeline}
\label{sec:binary}

The binary \ch{Nb-V} system is the first demonstration of the complete framework: an MLIP (ORB) supplies the thermodynamic database, Jansson interpolation evaluates chemical potentials and their composition derivatives directly from the CALPHAD dataset --- bypassing finite-difference approximations --- and the spectral Cahn-Hilliard solver propagates the resulting elasto-chemical driving forces on the grid. The single independent composition field isolates the Jansson derivative from multi-component complexity, and the well-characterized miscibility gap at \SI{500}{\celsius} with no spectator elements provides a clean validation setting before extension to the ternary and quaternary subsystems. The equilibrium BCC\textsubscript{A2} tie-line endpoints at \SI{500}{\celsius} are an Nb-rich phase ($x_{\rm Nb} = 0.96$) and a V-rich phase ($x_{\rm Nb} = 0.36$), between which the stiffness tensors $C_{ij}$ and Vegard eigenstrain coefficient $\alpha_{\rm Nb} = \partial\ln a/\partial x_{\rm Nb}$ are linearly interpolated from SQS configurations evaluated with the ORB MLIP. The coefficient $\alpha_{\rm Nb} = 0.0874$ is evaluated at the tie-line midpoint ($x_{\rm Nb} = 0.66$, $a_{\rm ref} = \SI{3.228}{\angstrom}$), giving a total misfit strain $\varepsilon_T = -5.5\%$ between the two equilibrium phases and a molar volume $N_V = \SI{8.99e4}{\mol\per\metre\cubed}$. \Cref{tab:binary_elastic} lists the full elastic constants and derived properties at both endpoints.

\begin{figure*}[!tp]
    \centering
    \includegraphics[width=\textwidth]{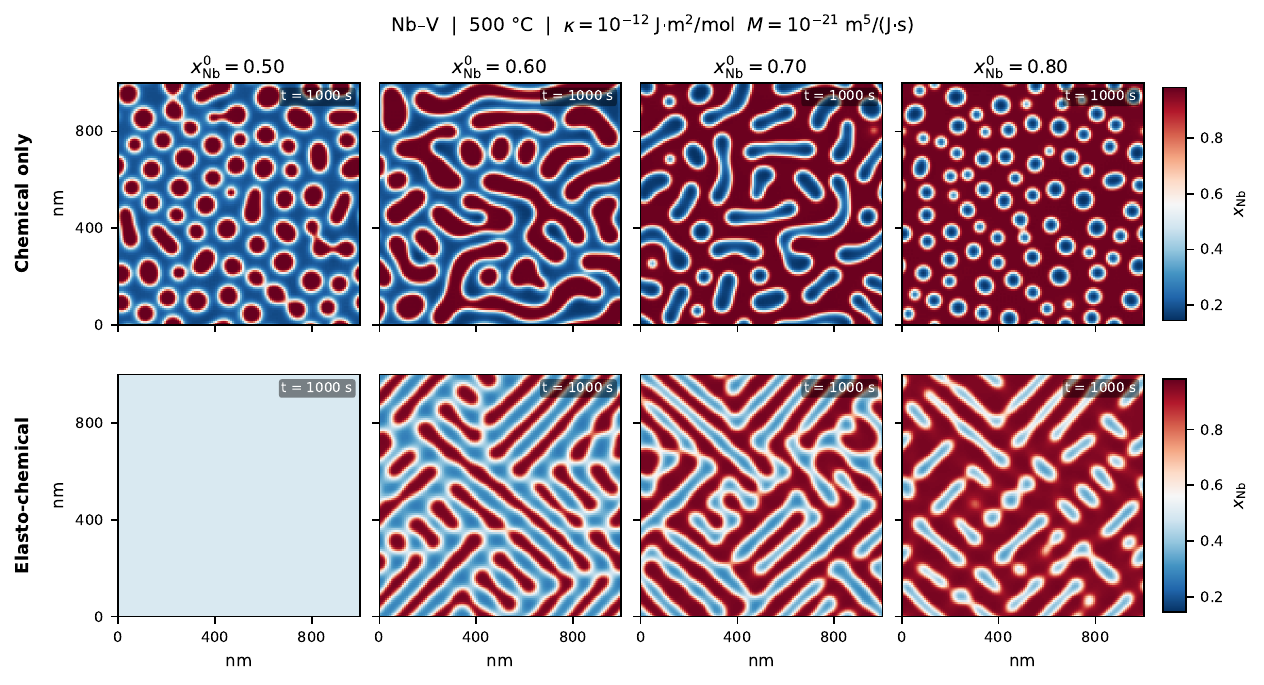}
    \caption{Simulated $x_{\rm Nb}$ microstructures from Cahn-Hilliard simulations of binary \ch{Nb-V} BCC\textsubscript{A2} at \SI{500}{\celsius} ($128\times128$ grid, $L = \SI{1}{\micro\metre}$, $\kappa = \SI{1e-12}{\joule\metre\squared\per\mol}$, $M = \SI{5e-21}{\metre\tothe{5}\per\joule\per\second}$). Top row: chemical only. Bottom row: elasto-chemical (tensorial Vegard coupling, $\alpha_{\rm Nb} = 0.087$, cubic stiffness from SQS/DFT). Columns correspond to mean compositions $x_{\rm Nb}^0 = 0.50,\,0.60,\,0.70,\,0.80$. Colormap: $x_{\rm Nb}$ increases from blue (low) to red (high). Both phases have $A_Z < 1$, selecting $\langle 110 \rangle$-aligned boundaries and the characteristic diamond-faceted morphology in the elasto-chemical case.}
    \label{fig:chem_vs_elastic}
\end{figure*}

Both endpoints have $A_Z < 1$ (\Cref{tab:binary_elastic}), identifying $\langle 110 \rangle$ as the elastically soft direction. \Cref{fig:chem_vs_elastic} shows the output of the MLIP-informed pipeline across four mean compositions. The chemical-only simulations (top row) produce isotropic interconnected domains at all four compositions, with no preferred orientation. The elasto-chemical simulations (bottom row) reveal a composition-dependent response that is absent in the chemical-only case. At $x_{\rm Nb}^0 = 0.50$, coherency strain completely suppresses spinodal decomposition and the field remains uniform, consistent with the coherent Hessian analysis showing that elastic energy closes the miscibility gap at the symmetric composition. At $x_{\rm Nb}^0 = 0.60$ and $0.70$, decomposition proceeds and produces a labyrinthine, bicontinuous morphology with channels preferentially aligned along $\langle 110 \rangle$, reflecting the elastically soft direction ($A_Z < 1$). At $x_{\rm Nb}^0 = 0.80$, the V-rich phase constitutes a small minority and forms isolated $\langle 110 \rangle$-elongated islands embedded in a continuous Nb-rich matrix.

The absence of decomposition at $x_{\rm Nb}^0 = 0.50$ reflects a direct thermodynamic consequence of coherency: at this composition and temperature, the elastic penalty is sufficient to render the smallest positive eigenvalue of the coherent Gibbs energy Hessian positive, stabilizing the alloy against all composition fluctuations. That the chemical-only simulation at the same composition decomposes freely establishes that this stability is entirely elastic in origin. Coherency strain thus closes the miscibility gap at the symmetric composition in \ch{Nb-V} --- an effect that a purely chemical phase field model would miss entirely.

\begin{table*}[!hbtp]
    \centering
    \caption{Elastic constants and Vegard eigenstrain parameters for the \ch{Nb-V} BCC\textsubscript{A2} tie-line endpoints at \SI{500}{\celsius}, computed from SQS configurations evaluated with the ORB machine-learned interatomic potential. $C' = (C_{11}-C_{12})/2$: tetragonal shear modulus; $A_Z = C_{44}/C'$: Zener anisotropy ratio; $\varepsilon_T = (a_{\rm ppt} - a_{\rm mat})/a_{\rm mat}$: misfit strain.}
    \label{tab:binary_elastic}
    \renewcommand{\arraystretch}{0.8}
    \setlength{\tabcolsep}{5pt}
    \begin{tabular}{lcc}
        \toprule
        Property & Nb-rich ($x_{\rm Nb}=0.96$) & V-rich ($x_{\rm Nb}=0.36$) \\
        \midrule
        \multicolumn{3}{l}{\textit{Composition and lattice parameter}} \\
        $x_{\rm Nb}$ & 0.96 & 0.36 \\
        $x_{\rm V}$  & 0.04 & 0.64 \\
        $a$ (\AA)    & 3.312 & 3.130 \\
        \midrule
        \multicolumn{3}{l}{\textit{Cubic elastic constants (GPa)}} \\
        $C_{11}$ & 243.6 & 259.3 \\
        $C_{12}$ & 119.7 & 145.0 \\
        $C_{44}$ &  20.6 &  18.0 \\
        $C' = (C_{11}-C_{12})/2$ & 62.0 & 57.2 \\
        \midrule
        \multicolumn{3}{l}{\textit{Derived elastic properties}} \\
        $G_{\rm VRH}$ (GPa) & 32.5 & 29.2 \\
        $B_{\rm VRH}$ (GPa) & 161.0 & 183.1 \\
        Poisson ratio $\nu$ & 0.406 & 0.424 \\
        Zener ratio $A_Z = C_{44}/C'$ & \textbf{0.332} & \textbf{0.315} \\
        \midrule
        \multicolumn{3}{l}{\textit{Vegard eigenstrain parameters (shared)}} \\
        $a_{\rm ref}$ (\AA) & \multicolumn{2}{c}{3.228} \\
        $\alpha_{\rm Nb} = \partial\ln a/\partial x_{\rm Nb}$ & \multicolumn{2}{c}{0.0874} \\
        $\varepsilon_T = \Delta a / a_{\rm mat}$ & \multicolumn{2}{c}{$-5.5\%$} \\
        $N_V$ (mol\,m$^{-3}$) & \multicolumn{2}{c}{$8.99\times10^{4}$} \\
        \bottomrule
    \end{tabular}
\end{table*}

\subsection{Ternary \ch{Nb-Hf-V}: Temperature and Composition Dependence}
\label{sec:ternary}

Hf functions as a dual-action tuning element in the \ch{Nb-V} spinodal: it deepens the chemical miscibility gap as its fraction increases, yet its large atomic radius simultaneously raises the coherency penalty that opposes decomposition. These two competing effects define a composition window — roughly $x_{\rm Hf}^0 \leq 0.25$ — within which spinodal microstructures are accessible under elasto-chemical coupling at \SI{500}{\celsius}, and place a strict upper bound on Hf content above which coherency strain closes the gap entirely.

The temperature dependence compounds this picture. \Cref{tab:spinodal_regimes} identifies four regimes for \ch{Nb_{50}Hf_5V_{45}} from the signs of the diagonal Hessian elements $H_{\rm Nb}$, $H_{\rm Hf}$, and $H_{\rm V} = H_{\rm Nb} + H_{\rm Hf} + 2H_{\rm Nb,Hf}$. Above \SI{700}{\celsius} no element is individually unstable. The spinodal survives only through weak off-diagonal \ch{Nb-Hf} coupling ($\lambda_{\min} \lesssim 0$) and produces no visible composition modulation. Between \SI{600}{\celsius} and \SI{700}{\celsius}, $H_{\rm Nb} < 0$ alone drives an asymmetric instability in which Nb segregates against a stable V$+$Hf background. Below \SI{500}{\celsius}, V also destabilizes ($H_{\rm V} < 0$), recovering binary-like \ch{Nb-V} demixing, but Hf then acts as a positive-curvature stiffener that severely constrains the stable time step (\Cref{eq:dt_stable_pf}). The optimal operating point is \SI{500}{\celsius}--\SI{700}{\celsius}, where the Nb-driven spinodal is thermodynamically active and Hf remains a spectator rather than a numerical bottleneck.

\begin{table*}[!tp]
    \centering
    \caption{Spinodal character and equilibrium two-phase compositions as a function of temperature for the binary \ch{Nb_{50}V_{50}} and ternary \ch{Nb_{50}Hf_5V_{45}} BCC\textsubscript{A2} alloys. Signs of the diagonal Hessian elements indicate whether each element is individually unstable ($H_{ii} < 0$, driving force) or stable ($H_{ii} > 0$, spectator). $H_{\rm V}$ is derived via $H_{\rm V} = H_{\rm Nb} + H_{\rm Hf} + 2H_{\rm Nb,Hf}$.}
    \label{tab:spinodal_regimes}
    \scriptsize
    \renewcommand{\arraystretch}{1.0}
    \setlength{\tabcolsep}{4pt}
    \newcommand{\unst}[1]{\textcolor{BrickRed}{$#1{<}0$}}
    \newcommand{\stab}[1]{\textcolor{gray}{$#1{>}0$}}
    \begin{tabularx}{\textwidth}{lp{3.0cm}lll>{\raggedright\arraybackslash}Xp{3.0cm}}
        \toprule
        System & Temperature & Instabilities & Hf role & $\Delta t$ & Character & BCC phases$^{a}$ \\
        \midrule
        Binary Nb$_{50}$V$_{50}$
            & $T < \SI{750}{\celsius}$
            & \unst{H_{\rm Nb}}\,=\,\unst{H_{\rm V}}
            & absent & fast
            & Symmetric Nb$\leftrightarrow$V demixing
            & BCC$_1$: Nb$_{99}$V$_{1}$; BCC$_2$: Nb$_1$V$_{99}$ \\
        \midrule
        \multirow{3}{*}{Ternary Nb$_{50}$Hf$_5$V$_{45}$}
            & $T > \SI{700}{\celsius}$
            & \stab{H_{\rm Nb}}, \stab{H_{\rm Hf}}, \stab{H_{\rm V}}
            & stable & fast
            & No instability; marginal off-diagonal \ch{Nb-Hf} coupling only
            & Single-phase BCC \\
            & \SIrange{600}{700}{\celsius}
            & \unst{H_{\rm Nb}}, \stab{H_{\rm Hf}}, \stab{H_{\rm V}}
            & spectator & moderate
            & Asymmetric Nb-driven spinodal; optimal window
            & BCC$_1$: Nb$_{90}$Hf$_9$V$_1$; BCC$_2$: Nb$_2$Hf$_{0.1}$V$_{98}$ \\
            & $T < \SI{500}{\celsius}$
            & \unst{H_{\rm Nb}}, \stab{H_{\rm Hf}}, \unst{H_{\rm V}}
            & stiffener & penalized
            & Binary-like \ch{Nb-V} demixing; $H_{\rm Hf}{>}0$ constrains $\Delta t$
            & BCC$_1$: Nb$_{91}$Hf$_9$V$_{0.3}$; BCC$_2$: Nb$_{0.6}$Hf$_{0.04}$V$_{99}$ \\
        \bottomrule
        \multicolumn{7}{l}{$^{a}$~Tie-line endpoints from \texttt{PyCalphad} (\texttt{BCC\_A2}, ORB-MLIP database); subscripts at.\%. \SI{500}{\celsius} for $T{<}\SI{750}{\celsius}$ rows, \SI{700}{\celsius} for \SIrange{600}{700}{\celsius} row.}
    \end{tabularx}
\end{table*}

\begin{figure*}[!hbtp]
    \centering
    \begin{overpic}[width=\textwidth]{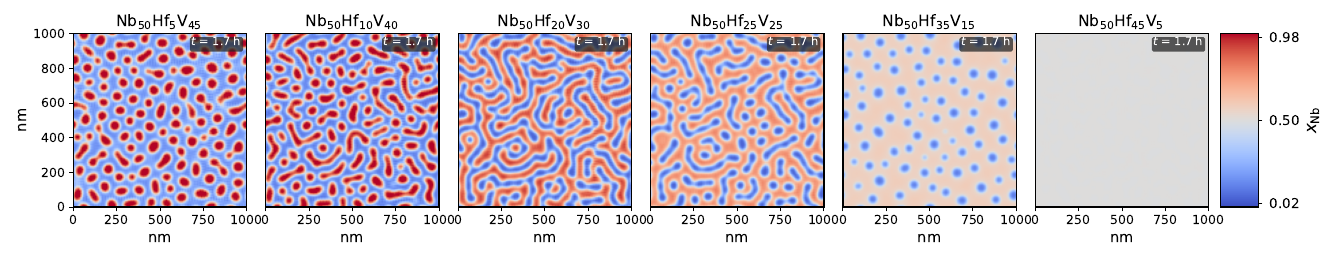}
    \put(0, 20){\textbf{(a)}}
    \end{overpic}
    %%%
    \begin{overpic}[width=\textwidth]{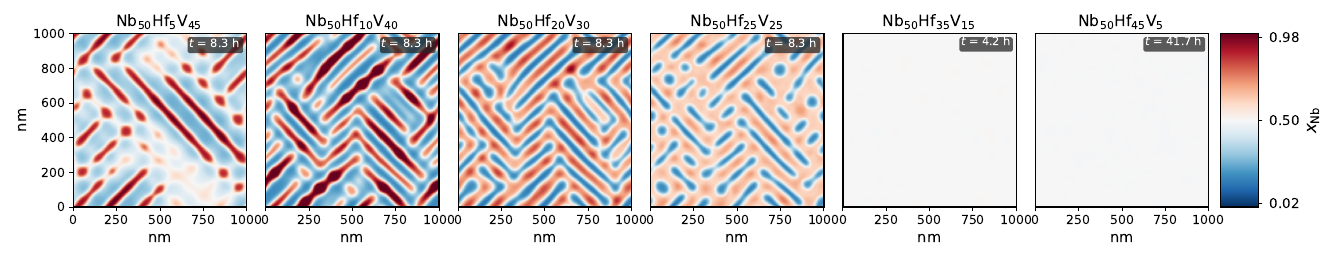}
    \put(0, 20){\textbf{(b)}}
    \end{overpic}
    %%%
    \begin{overpic}[width=0.6\textwidth]{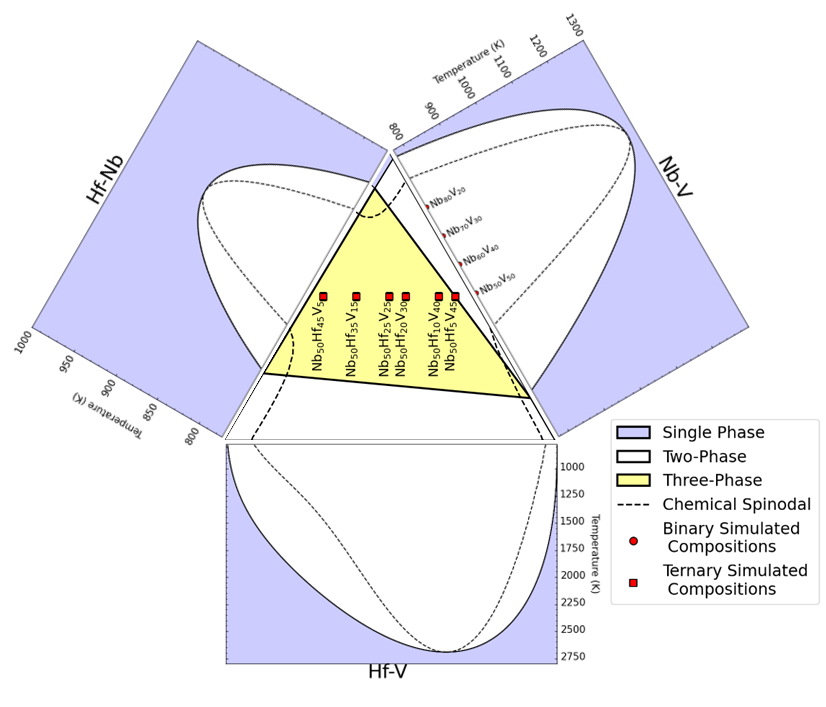}
    \put(0, 80){\textbf{(c)}}
    \end{overpic}
    \caption{Hf-content sweep of $x_{\rm Nb}$ microstructures in ternary \ch{Nb-Hf-V} BCC\textsubscript{A2} at \SI{500}{\celsius} ($128\times128$ grid, $L = \SI{1}{\micro\metre}$, $\kappa = \SI{1e-12}{\joule\metre\squared\per\mol}$, $M_{\rm Nb} = \SI{5e-21}{\metre\tothe{5}\per\joule\per\second}$). \textbf{(a)} Chemical-only simulations, all compositions after \SI{6000}{\second} of isothermal aging. \textbf{(b)} Elasto-chemical simulations using the composition-dependent coherency eigenstrain strain model; snapshots after \SI{30000}{\second} (Nb$_{50}$Hf$_{5}$V$_{45}$, Nb$_{50}$Hf$_{10}$V$_{40}$, Nb$_{50}$Hf$_{20}$V$_{30}$, Nb$_{50}$Hf$_{25}$V$_{25}$), \SI{15000}{\second} (Nb$_{50}$Hf$_{35}$V$_{15}$), and \SI{150000}{\second} (Nb$_{50}$Hf$_{45}$V$_{5}$). Compositions with $x_{\rm Hf}^0 \geq 0.35$ show no spinodal decomposition, demonstrating that coherency strain suppresses the miscibility gap at high Hf content. \textbf{(c)} Ternary BCC\textsubscript{A2} miscibility gap across all three binary subsystems (\ch{Hf-Nb}, \ch{Nb-V}, \ch{Hf-V}). Colormap: $x_{\rm Nb}$ increases from blue (low) to red (high).}
    \label{fig:ternary_summary}
\end{figure*}

\Cref{fig:ternary_summary} consolidates three further perspectives on the ternary system. \Cref{fig:ternary_summary}(a) shows the chemical-only Hf-content sweep across six compositions ($x_{\rm Hf}^0 = 0.05$--$0.45$), each after \SI{6000}{\second} of isothermal aging at \SI{500}{\celsius}. At low Hf fractions ($x_{\rm Hf}^0 = 0.05$, 0.10) decomposition produces isolated Nb-rich precipitates consistent with a minority-phase microstructure. Increasing Hf to 0.20 and 0.25 shifts the tie-line endpoints and produces well-developed bicontinuous \ch{Nb-V} domains, indicating that Hf progressively deepens the miscibility gap. At $x_{\rm Hf}^0 = 0.35$ the domains coarsen to a lower-contrast pattern, and in \ch{Nb_{50}Hf_{45}V_{5}} no decomposition occurs, placing this composition outside the chemical spinodal at \SI{500}{\celsius}.

\Cref{fig:ternary_summary}(b) shows the corresponding elasto-chemical sweep using the composition-dependent coherency eigenstrain model; snapshots are taken after \SI{30000}{\second} for \ch{Nb_{50}Hf_{5}V_{45}}, \ch{Nb_{50}Hf_{10}V_{40}}, \ch{Nb_{50}Hf_{20}V_{30}}, and \ch{Nb_{50}Hf_{25}V_{25}}; \SI{15000}{\second} for \ch{Nb_{50}Hf_{35}V_{15}}; and \SI{150000}{\second} for \ch{Nb_{50}Hf_{45}V_{5}}. The elastic coupling further narrows the spinodal region: \ch{Nb_{50}Hf_{5}V_{45}}, \ch{Nb_{50}Hf_{10}V_{40}}, \ch{Nb_{50}Hf_{20}V_{30}}, and \ch{Nb_{50}Hf_{25}V_{25}} still decompose with phase contrast and domain connectivity similar to the chemical-only case, while \ch{Nb_{50}Hf_{35}V_{15}} and \ch{Nb_{50}Hf_{45}V_{5}} remain uniform throughout the simulation. The suppression at $x_{\rm Hf}^0 = 0.35$ is purely elastic in origin: the chemical-only simulation at the same composition decomposes, whereas coherency strain raises the effective free-energy curvature above zero and stabilizes the solid solution. This trend — elastic stabilization becoming dominant as Hf content increases — is consistent with the larger Vegard misfit parameter of Hf relative to V and the attendant increase in elastic self-energy at Hf-rich compositions. \Cref{fig:ternary_summary}(c) maps the simulated compositions onto the ternary BCC\textsubscript{A2} miscibility gap, projected across the \ch{Hf-Nb}, \ch{Nb-V}, and \ch{Hf-V} binary subsystems.

\subsection{Quaternary \ch{Hf-Nb-Ti-V}: Phase Field Prediction and Experimental Validation}
\label{sec:quaternary}

\begin{figure*}[!tp]
    \centering
    \begin{overpic}[width=0.32\textwidth]{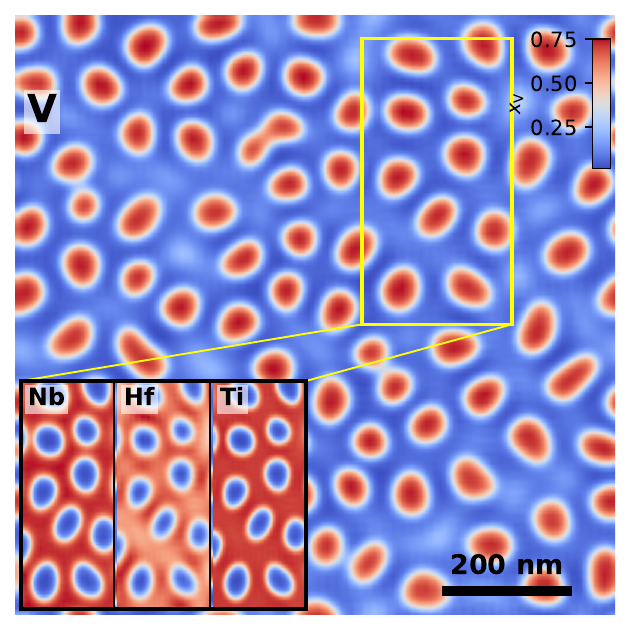}
        \put(3,92){\colorbox{black!25}{\makebox(6.1,3.4){\textbf{(a)}}}}
    \end{overpic}\hfill
    \begin{overpic}[width=0.32\textwidth]{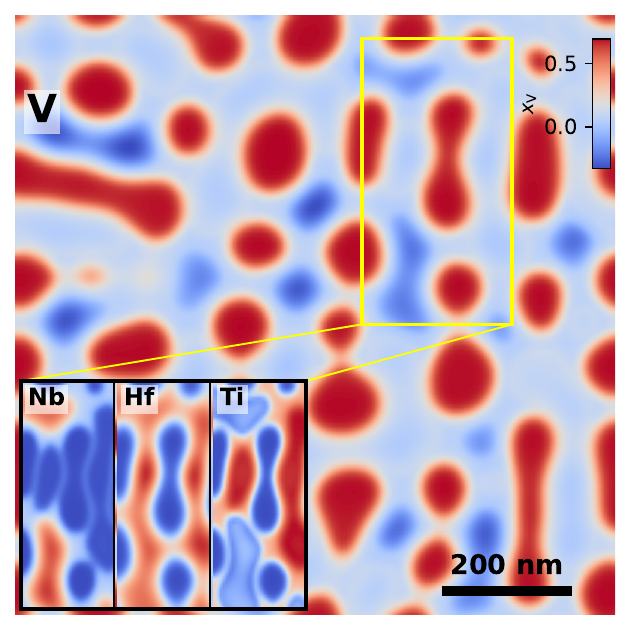}
        \put(3,92){\colorbox{black!25}{\makebox(6.1,3.4){\textbf{(b)}}}}
    \end{overpic}\hfill
    \begin{overpic}[width=0.345\textwidth]{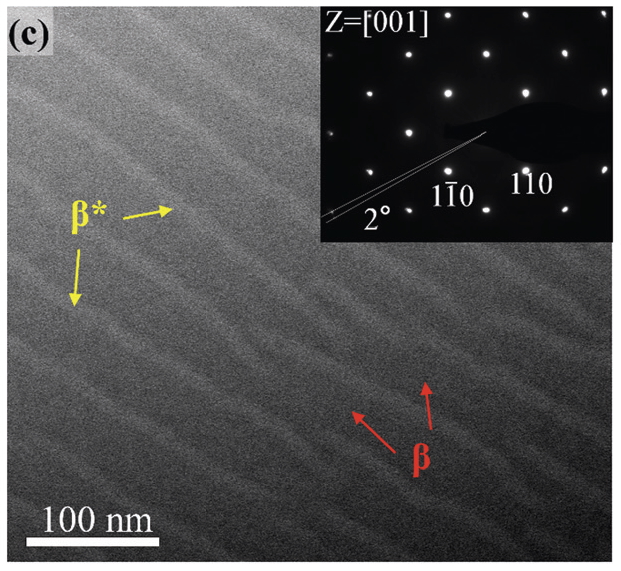}
        %\put(1,86){\colorbox{black!25}{\makebox(6.1,3.4){\textbf{(c)}}}}
    \end{overpic}
    \caption{Phase field simulations of equiatomic \ch{Nb_{25}Hf_{25}Ti_{25}V_{25}} at \SI{500}{\celsius} and HAADF-STEM comparison. (a)~Chemical-only simulation at \SI{60000}{\second}: $x_{\rm V}$ field ($128\times128$, $L = \SI{1}{\micro\metre}$) with insets showing $x_{\rm Nb}$, $x_{\rm Hf}$, $x_{\rm Ti}$ for the highlighted region. (b)~Elasto-chemical simulation at \SI{700000}{\second}: same fields with elastic coupling. (c)~HAADF-STEM image of as-cast alloy showing alternating $\beta$/$\beta^*$ BCC\textsubscript{A2} variants; reproduced from An \etal \cite{an2021spinodal}.}
    \label{fig:quaternary_chem}
\end{figure*}

Equiatomic \ch{Nb_{25}Hf_{25}Ti_{25}V_{25}} lies within the BCC\textsubscript{A2} spinodal at \SI{500}{\celsius} ($\lambda_{\min} < 0$) and produces well-evolved microstructure on accessible simulation timescales. The chemical-only simulation at \SI{60000}{\second} yields ellipsoidal V-rich precipitates ($x_{\rm V}^{\rm max} = 0.751$) in an Nb/Hf/Ti-rich matrix (\Cref{fig:quaternary_chem}(a)). Diffuse Hf halos around precipitate cores arise from the tenfold lower Hf mobility rather than an independent Hf instability, consistent with the kinetic analysis in \Cref{sec:ternary}.

The elasto-chemical simulation at \SI{700000}{\second} reproduces the same V-rich/Nb-Ti-rich partitioning but with reduced compositional contrast ($x_{\rm V}^{\rm max} = 0.691$), reflecting the elastic penalty that opposes sharp composition gradients (\Cref{fig:quaternary_chem}(b)). CALPHAD equilibrium predicts two coexisting BCC\textsubscript{A2} phases ($f_{\rm ppt} \approx 0.33$, \Cref{tab:quaternary_tieline}). Both simulations approach the CALPHAD endpoints for Nb ($x_{\rm Nb}^{\rm mat} \approx 0.30$) and V ($x_{\rm V}^{\rm ppt} \approx 0.71$), while Hf and Ti remain far from their equilibrium values, consistent with slow Hf kinetics and the limited simulation timescales accessible.

\begin{table*}[!tp]
    \centering
    \caption{CALPHAD tie-line endpoints for equiatomic \ch{Nb_{25}Hf_{25}Ti_{25}V_{25}} BCC\textsubscript{A2} at \SI{500}{\celsius} (ORB-MLIP database), with chemical and elasto-chemical phase field composition ranges shown for comparison. PFM values are non-equilibrium, particularly for Hf and V.}
    \label{tab:quaternary_tieline}
    \scriptsize
    \renewcommand{\arraystretch}{1.0}
    \setlength{\tabcolsep}{5pt}
    \begin{tabular}{llcccc}
        \toprule
        & & $x_{\rm Nb}$ & $x_{\rm Hf}$ & $x_{\rm Ti}$ & $x_{\rm V}$ \\
        \midrule
        \multirow{2}{*}{CALPHAD tie-line}
            & Matrix ($f \approx 0.67$)
            & 0.300 & 0.371 & 0.305 & 0.024 \\
            & Precipitate ($f \approx 0.33$)
            & 0.148 & 0.002 & 0.136 & 0.714 \\
        \midrule
        \multirow{2}{*}{Chemical PFM (\SI{60000}{\second})}
            & Matrix (max)
            & 0.331 & 0.365 & 0.310 & 0.751 \\
            & Precipitate (min)
            & 0.096 & 0.048 & 0.103 & 0.015 \\
        \midrule
        \multirow{2}{*}{Elasto-chem.\ PFM (\SI{700000}{\second})}
            & Matrix (max)
            & 0.396 & 0.426 & 0.353 & 0.691 \\
            & Precipitate (min)
            & 0.127 & 0.056 & 0.125 & 0.053 \\
        \bottomrule
    \end{tabular}
\end{table*}

The MLIP-computed elastic constants predict a spinodal modulation direction in the quaternary that differs qualitatively from the binary and ternary baselines: the SQS/ORB-MLIP interpolant evaluated at the CALPHAD matrix composition yields $A_Z^{\rm mat} = 1.52 > 1$, selecting $\langle 100 \rangle$ as the elastically soft direction and therefore the preferred microstructure alignment, opposite to the binary \ch{Nb-V} ($A_Z = 0.33 < 1$) and ternary \ch{Nb-Hf-V} ($A_Z = 0.37 < 1$) cases where $\langle 110 \rangle$ is favored. \Cref{tab:elastic_comparison} compares the full set of model predictions against the experimental characterization of equiatomic \ch{Hf-Nb-Ti-V} by An \etal \cite{an2021spinodal}, who identified two coexisting BCC\textsubscript{A2} phases in an as-cast alloy by XRD, HAADF-STEM, and EDS, revealing three areas of agreement and disagreement.

\begin{table*}[!tp]
    \centering
    \caption{Predicted and experimentally measured properties for the equiatomic BCC\textsubscript{A2} \ch{Hf-Nb-Ti-V} alloy. Experimental values are from An \etal \cite{an2021spinodal} (as-cast, characterized by XRD, HAADF-STEM, and EDS). Binary predictions use the Nb$_{50}$V$_{50}$ CALPHAD tie-line endpoints (matrix: Nb$_{96}$V$_{4}$; precipitate: Nb$_{36}$V$_{64}$) at \SI{500}{\celsius}. Ternary predictions use the Nb$_{50}$Hf$_{25}$V$_{25}$ CALPHAD tie-line endpoints (matrix: Nb$_{70}$Hf$_{10}$V$_{20}$; precipitate: Nb$_{15}$Hf$_{35}$V$_{50}$) at \SI{500}{\celsius}. Quaternary predictions use the equiatomic Nb$_{25}$Hf$_{25}$Ti$_{25}$V$_{25}$ CALPHAD tie-line at \SI{500}{\celsius} (\Cref{tab:quaternary_tieline}). $A_Z = 2C_{44}/(C_{11}-C_{12})$: $A_Z > 1$ selects $\langle 100 \rangle$ modulation, $A_Z < 1$ selects $\langle 110 \rangle$. $\varepsilon_T = (a_{\rm ppt} - a_{\rm mat})/a_{\rm mat}$.}
    \label{tab:elastic_comparison}
    \scriptsize
    \renewcommand{\arraystretch}{1.0}
    \setlength{\tabcolsep}{4pt}
    \begin{tabularx}{\textwidth}{lXXXX}
        \toprule
        Property
            & Binary Nb$_{50}$V$_{50}$ (this work)
            & Ternary Nb$_{50}$Hf$_{25}$V$_{25}$ (this work)
            & Quaternary Nb$_{25}$Hf$_{25}$Ti$_{25}$V$_{25}$ (this work)
            & Experiment \cite{an2021spinodal} \\
        \midrule
        Precipitate enriched in
            & V
            & V, Hf
            & V (Hf in matrix)
            & Hf $+$ V \\
        Matrix enriched in
            & Nb
            & Nb (Hf depleted)
            & Nb, Hf, Ti
            & Nb $+$ Ti \\
        Precipitate phase fraction
            & ---
            & ---
            & $\approx 0.33$
            & $\approx 1/3$ \\
        $a_{\rm mat}$ (\AA)
            & 3.312
            & 3.286
            & 3.382
            & 3.313 \\
        $a_{\rm ppt}$ (\AA)
            & 3.130
            & 3.265
            & 3.089
            & 3.259 \\
        $\varepsilon_T$
            & $-5.5\%$
            & $-0.64\%$
            & $-8.7\%$
            & $-1.6\%$ (XRD); $0.54\%$ (TEM) \\
        $A_Z$ (matrix)
            & 0.33
            & 0.37
            & \textbf{1.52}
            & --- \\
        $A_Z$ (precipitate)
            & 0.31
            & 1.09
            & 0.32
            & --- \\
        Spinodal modulation direction
            & $\langle 110 \rangle$ (predicted)
            & $\langle 110 \rangle$ (predicted)
            & $\langle 100 \rangle$ (predicted)
            & $\langle 110 \rangle$ (observed) \\
        Modulation wavelength
            & ---
            & ---
            & ${\approx}50$--\SI{100}{\nano\metre}
            & 50--\SI{100}{\nano\metre} \\
        \bottomrule
    \end{tabularx}
\end{table*}

Hf partitioning is reversed between the CALPHAD model and experiment. An \etal demonstrate via HAADF-STEM Z-contrast that Hf ($Z = 72$, the heaviest constituent) enriches the precipitate phase ($\beta^*$), co-segregating with V. The ORB-MLIP CALPHAD model assigns Hf to the matrix alongside Nb and Ti (\Cref{tab:quaternary_tieline}). The most probable explanation is non-equilibrium partitioning during rapid solidification: the experimental alloy was drop-cast, so the system may not have reached the \SI{500}{\celsius} equilibrium partitioning predicted by CALPHAD, and instead retains a composition field set during or shortly after solidification. Database inaccuracies in the ORB-MLIP model for heavy-element BCC alloys cannot be excluded. The low equilibrium diffusivity of Hf supports the view that the system has not yet relaxed to thermodynamic equilibrium, making non-equilibrium trapping of the as-cast partitioning plausible.

The coherency misfit is overestimated, directly tracing to the Hf partitioning error. An \etal report lattice misfits of $1.6\%$ (XRD peak separation) and $0.54\%$ (TEM atomic-column measurements), both substantially smaller in magnitude than the $8.7\%$ predicted by the quaternary CALPHAD tie-line. If Hf (large atomic radius, $a_{\rm Hf}^{\rm BCC} \approx \SI{3.62}{\angstrom}$) co-partitions with V (small radius, $a_{\rm V}^{\rm BCC} \approx \SI{3.03}{\angstrom}$) into the precipitate, their opposing contributions to the precipitate lattice parameter partially cancel, yielding a net misfit consistent with experiment. This internal consistency supports the thermal-history origin of the discrepancy. The binary \ch{Nb-V} misfit of $-5.5\%$ (\Cref{tab:elastic_comparison}) is similarly overestimated relative to available experimental \ch{Nb-V} lattice data~\cite{pearson2013handbook,de2015charting}, though less severely.

\begin{figure}[!tp]
    \centering
    \includegraphics[width=\linewidth]{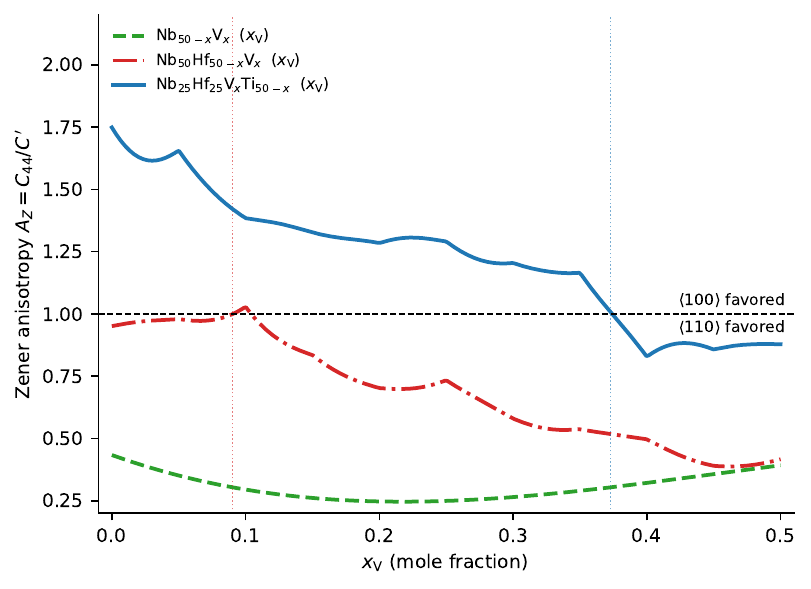}
    \caption{Zener anisotropy $A_Z = C_{44}/C'$ from SQS/ORB-v2 MLIP elastic constants along three composition paths plotted against $x_{\rm V}$ (mole fraction). Binary Nb$_{50-x}$V$_{x}$ (green dashed): $A_Z < 1$ throughout, $\langle 110 \rangle$ favored. Ternary Nb$_{50}$Hf$_{50-x}$V$_{x}$ (red dash-dot): $A_Z$ briefly exceeds unity near $x_{\rm V} \approx 0.10$ before falling below 1. Quaternary Nb$_{25}$Hf$_{25}$V$_{x}$Ti$_{50-x}$ (blue solid): as V replaces Ti, $A_Z$ crosses unity at $x_{\rm V} \approx 0.37$ (equivalently $x_{\rm Ti} \approx 0.13$), switching the predicted modulation direction from $\langle 100 \rangle$ to $\langle 110 \rangle$. Dotted vertical lines mark the $A_Z = 1$ crossover compositions.}
    \label{fig:az_vs_ti}
\end{figure}

The quaternary elastic anisotropy does not correctly predict the observed $\langle 110 \rangle$ spinodal modulation direction. The SAED pattern in An \etal (Fig.~1(c) inset, $[001]$ zone axis) shows diffraction spots along $\langle 110 \rangle$, and the $\beta^*$ lamellae visible in the HAADF-STEM image are elongated along $\langle 110 \rangle$, identifying this as the experimental modulation direction. The quaternary CALPHAD model gives $A_Z^{\rm mat} = 1.52 > 1$, which predicts $\langle 100 \rangle$ modulation, inconsistent with the observed $\langle 110 \rangle$. The binary \ch{Nb-V} and ternary \ch{Nb-Hf-V} models ($A_Z < 1$) also predict $\langle 110 \rangle$, but this agreement is coincidental: these compositions lie in a V-rich regime where $A_Z < 1$ regardless of Ti content, and they do not capture the elastic behavior of the actual quaternary matrix. Interpolating the SQS/ORB-MLIP elastic constants along a compositional path from the ternary matrix (\ch{Nb_{70}Hf_{10}V_{20}}, $A_Z = 0.37$) to the quaternary matrix (\ch{Nb_{30}Hf_{37}Ti_{31}V_{2}}, $A_Z = 1.52$) reveals that $A_Z$ crosses unity at $x_{\rm V} \approx 0.37$ (equivalently $x_{\rm Ti} \approx 0.13$), identifying Ti addition as the element responsible for flipping the predicted modulation direction from $\langle 110 \rangle$ to $\langle 100 \rangle$ (\Cref{fig:az_vs_ti}). Ti in the BCC phase exhibits a high $C_{44}/C'$ ratio that preferentially stiffens the relevant shear mode, driving $A_Z$ above unity as its concentration increases~\cite{soderlind1993theory}. Note that BCC Ti is metastable at ambient conditions and requires DFT or MLIP calculations to characterize its elastic tensor. This discrepancy may reflect the as-cast, non-equilibrium character of the experimental microstructure, where kinetics and thermal history rather than the equilibrium elastic anisotropy govern the early-stage modulation direction. Despite these discrepancies, the model correctly predicts a phase fraction of $\approx 1/3$ and a modulation wavelength of ${\approx}50$--\SI{100}{\nano\metre}, both in close agreement with the experimental observations of An \etal.

The quantitative discrepancies observed in the quaternary system point primarily to limitations of the current MLIP description. The SQS/ORB-MLIP elastic constants are 0~K static estimates that have not been validated against DFT or experimental measurements for the specific quaternary compositions predicted by the CALPHAD tie-line. Errors in these values propagate directly into both the coherency misfit and the $A_Z$ prediction. Moreover, the 0~K static approximation does not capture finite-temperature lattice softening or the effect of short-range chemical order on the shear moduli. Both effects are significant in BCC refractory alloys at elevated temperatures~\cite{ikeda2019ab}. Non-equilibrium solidification in the as-cast specimen may contribute to the Hf partitioning discrepancy, but the systematic nature of the elastic errors suggests that dedicated DFT validation of the quaternary elastic constants and an improved finite-temperature MLIP description are the most impactful paths toward quantitative agreement with experiment.
\section{Conclusions and Future Work}

In this work, we demonstrated (1) an open-source workflow for high-throughput microstructure stability analysis and (2) an MLIP-to-microstructure prediction pipeline. If a suitable CALPHAD model is not available to the modeler, it can be created through PhaseForge. The chemical Hessian is then determined analytically and efficiently via Jansson derivatives. Coherent strain contributions are added through an analytically differentiable bond model and MLIP estimates of the elastic constants. Finally, following a change of basis, gauge-invariant eigenvalues and eigenvectors of the chemical and coherent spinodal can be investigated at specific compositions as functions of temperature, or visualized at a given temperature across a high-dimensional composition space through affine projections. Together, this workflow extends the foundational stability analysis framework of Koneru, Kadirvel, and Wang \cite{koneru2022high} in five ways: (1) open-source access, (2) whitebox CALPHAD model knowledge, (3) improved efficiency and robustness of Gibbs energy derivatives, (4) modeling coherent strain contributions, and (5) visualization in large composition spaces via a single two-dimensional plot.

A further extension over prior work is the prediction of microstructure evolution and morphology, obtained by feeding the PhaseForge-generated thermodynamic model and calculated elastic constants into a CALPHAD-coupled elasto-chemical phase field simulation. Within these simulations, Jansson-derivative surrogates for the Gibbs energy Hessian ensure accurate length-scale predictions by eliminating the truncation errors of finite-difference approximations.

We applied the framework to the BCC phase of the \ch{Hf-Nb-Ti-V} system, progressing from binary \ch{Nb-V} through ternary \ch{Nb-Hf-V} to the quaternary. Coherency contributions systematically shrink the spinodal region: they close the miscibility gap entirely at the symmetric \ch{Nb-V} composition, and confine spinodal-accessible ternary alloys to $x_{\rm Hf}\lesssim 0.25$. Hf functions as a dual-action tuning element that deepens the chemical driving force while simultaneously raising the coherency penalty. Quaternary phase field predictions for equiatomic \ch{Nb_{25}Hf_{25}Ti_{25}V_{25}} reproduce the phase fraction ($\approx 1/3$), modulation wavelength (50--\SI{100}{\nano\metre}), and bicontinuous morphology reported by An~\etal~\cite{an2021spinodal}. Three quantitative discrepancies remain: Hf partitions to the matrix rather than the precipitate, the coherency misfit is overestimated ($8.7\%$ vs.\ $1.6\%$ XRD / $0.54\%$ TEM), and the predicted $\langle 100 \rangle$ modulation direction disagrees with the observed $\langle 110 \rangle$. The partitioning reversal is consistent with non-equilibrium as-cast solidification, while the elastic-anisotropy discrepancy traces to the 0~K static SQS/MLIP elastic constants and the Ti-driven inversion of the Zener ratio at Ti-rich matrix compositions.

Several future directions can extend this workflow. Fine-tuning the MLIP to the material system of interest could improve CALPHAD model and mechanical property predictions, particularly for mechanically unstable reference structures such as BCC Hf. For systems where experimental or computational thermodynamic data are available, gradient-based optimization schemes can efficiently refine the CALPHAD model \cite{kunselman2025construction}. Direct DFT validation of the quaternary elastic constants would clarify whether the Zener-ratio inversion observed at Ti-rich matrix compositions is a physical effect or an MLIP artifact. Temperature dependence could be introduced into the elastic constants through molecular dynamics: while MD is too expensive to cover a fine grid over a high-dimensional composition space, it can be used to build a surrogate for temperature-dependent corrections added to the temperature-independent MLIP-calculated values, addressing the finite-temperature lattice softening known to be significant in BCC refractory alloys. Finally, implementation of second-order Jansson derivatives would provide analytic calculation of third-order derivatives of the Gibbs energy with respect to composition, enabling the analytic calculation of critical points in high-dimensional design spaces.

\section*{Code Availability}

The thermodynamic databases used in this work were generated using PhaseForge~\cite{sariturk_2025_15730911} (\url{https://doi.org/10.5281/zenodo.15730911}). Elastic constants were computed from SQS configurations evaluated with ORB-v2 using MaterialsFramework~\cite{sariturk_2025_15731044} (\url{https://doi.org/10.5281/zenodo.15731044}). The CALPHAD-coupled phase-field solver, spinodal stability analysis workflow, and all simulation configs and post-processing scripts are openly available at \url{https://github.com/vahid2364/PhaseFoundry}.
\section*{Acknowledgments}
 RA also wishes to acknowledge the support from NSF through Grant No. NSF-DMREF-2119103. R.A.~and D.S.~acknowledge support from the Army Research Office (ARO) through Grant No.~W911NF-22-2-0117. Portions of this research were conducted with the advanced computing resources provided by Texas A\&M High Performance Research Computing.

\bibliographystyle{elsarticle-num}
\biboptions{numbers,sort&compress}
\bibliography{refs,refs_Vahid}

@article{gibbs1878equilibrium,
  title={On the equilibrium of heterogeneous substances},
  author={Gibbs, Josiah Willard},
  journal={American Journal of Science},
  volume={3},
  number={96},
  pages={441--458},
  year={1878},
}

@article{cahn1961spinodal,
  title={On spinodal decomposition},
  author={Cahn, John W},
  journal={Acta Metallurgica},
  volume={9},
  number={9},
  pages={795--801},
  year={1961},
}

@article{singh2015atomic,
  title={Atomic short-range order and incipient long-range order in high-entropy alloys},
  author={Singh, Prashant and Smirnov, Andrei V and Johnson, Duane D},
  journal={Physical Review B},
  volume={91},
  number={22},
  pages={224204},
  year={2015},
}

@article{de1972analysis,
  title={{An analysis of clustering and ordering in multicomponent solid solutions-I. Stability criteria}},
  author={De Fontaine, D},
  journal={Journal of Physics and Chemistry of Solids},
  volume={33},
  number={2},
  pages={297--310},
  year={1972},
}

@phdthesis{de1967computer,
  title={{A computer simulation of the evolution of coherent composition variations in solid solutions}},
  author={De Fontaine, Didier},
  year={1967},
  school={Northwestern University}
}

@article{kadirvel2022exploration,
  title={{Exploration of spinodal decomposition in multi-principal element alloys (MPEAs) using CALPHAD modeling}},
  author={Kadirvel, Kamalnath and Koneru, Shalini Roy and Wang, Yunzhi},
  journal={Scripta Materialia},
  volume={214},
  pages={114657},
  year={2022},
}

@incollection{otis2017pycalphad,
  title={{pycalphad: CALPHAD-based Computational Thermodynamics in Python}},
  author={Otis, Richard and Liu, Zi-Kui},
  booktitle={Zentropy},
  pages={373--392},
  year={2017},
  publisher={Jenny Stanford Publishing}
}

@article{kunselman2024analytically,
  title={Analytically differentiable metrics for phase stability},
  author={Kunselman, Courtney and Bocklund, Brandon and van de Walle, Axel and Otis, Richard and Arr{\'o}yave, Raymundo},
  journal={Calphad},
  volume={86},
  pages={102705},
  year={2024},
}

@article{morral2021stability,
  title={Stability of high entropy alloys to spinodal decomposition},
  author={Morral, John E and Chen, Shuanglin},
  journal={Journal of Phase Equilibria and Diffusion},
  volume={42},
  number={5},
  pages={673--695},
  year={2021},
}

@article{wang2023spinodal,
  title={{Spinodal decomposition induced brittleness of Zr-Ta containing medium-entropy alloys}},
  author={Wang, Shubin and Wang, Junfeng and Shu, Da and Shi, Peiying and Wu, Mingxu and Wang, Donghong and Yang, Chao and Zhu, Guoliang and Sun, Baode},
  journal={Materials Characterization},
  volume={205},
  pages={113330},
  year={2023},
}

@article{dasari2023crystallographic,
  title={{Crystallographic and compositional evolution of ordered B2 and disordered BCC phases during isothermal annealing of refractory high-entropy alloys}},
  author={Dasari, Sriswaroop and Sharma, Abhishek and Soni, Vishal and Kloenne, Zachary and Fraser, Hamish and Banerjee, Rajarshi},
  journal={Microscopy and Microanalysis},
  volume={29},
  number={1},
  pages={303--313},
  year={2023},
}

@article{an2021spinodal,
  title={Spinodal-modulated solid solution delivers a strong and ductile refractory high-entropy alloy},
  author={An, Zibing and Mao, Shengcheng and Yang, Tao and Liu, Chain Tsuan and Zhang, Bin and Ma, Evan and Zhou, Hao and Zhang, Ze and Wang, Lihua and Han, Xiaodong},
  journal={Materials Horizons},
  volume={8},
  number={3},
  pages={948--955},
  year={2021},
}

@article{lee2025spinodal,
  title={{Spinodal-assisted multiscale immiscibility enables superior strength--ductility synergy in (CuFeMnNi) 96Al2Ti2 high-entropy alloy}},
  author={Lee, Shi Woo and Park, Hyojin and Kim, Rae Eon and Kim, Jaehun and Lee, Do Won and Lee, Jae Heung and Hong, Sun Ig and Jang, Hyo-Sun and Heo, Yoon-Uk and Kim, Hyoung Seop},
  journal={Materials Science and Engineering: A},
  pages={149582},
  year={2025},
}

@article{zhang2022refinement,
  title={Refinement strengthening, second phase strengthening and spinodal microstructure-induced strength-ductility trade-off in a high-entropy alloy},
  author={Zhang, Wei and Ma, Zhichao and Zhao, Hongwei and Ren, Luquan},
  journal={Materials Science and Engineering: A},
  volume={847},
  pages={143343},
  year={2022},
}

@article{dong2024cooperative,
  title={Cooperative regulation of mechanical properties and magnetoresistance effect in high-entropy alloys by spinodal decomposition},
  author={Dong, Peilin and Zhang, Lei and Huang, Liufei and Yang, Qiuju and Li, Lin and Ma, Lei and Zhong, Zhiyong and Li, Jinfeng},
  journal={Journal of Alloys and Compounds},
  volume={970},
  pages={172547},
  year={2024},
}

@article{duan2025spinodal,
  title={{Spinodal decomposition promoting high thermoelectric performance in half-Heusler}},
  author={Duan, Sichen and Bao, Xin and Huang, Jiawei and Shi, Rongpei and Fei, Linfeng and Xue, Wenhua and Yao, Honghao and Li, Xiaofang and Wang, Jian and Liu, Xingjun and others},
  journal={Joule},
  volume={9},
  number={4},
  year={2025},
}

@article{koneru2022high,
  title={High-throughput design of multi-principal element alloys with spinodal decomposition assisted microstructures},
  author={Koneru, Shalini Roy and Kadirvel, Kamalnath and Wang, Yunzhi},
  journal={Journal of Phase Equilibria and Diffusion},
  volume={43},
  number={6},
  pages={753--763},
  year={2022},
}

@article{usanmaz2016first,
  title={First principles thermodynamical modeling of the binodal and spinodal curves in lead chalcogenides},
  author={Usanmaz, Demet and Nath, Pinku and Plata, Jose J and Hart, Gus LW and Takeuchi, Ichiro and Nardelli, Marco Buongiorno and Fornari, Marco and Curtarolo, Stefano},
  journal={Physical Chemistry Chemical Physics},
  volume={18},
  number={6},
  pages={5005--5011},
  year={2016},
}

@article{divilov2024priori,
  title={A priori procedure to establish spinodal decomposition in alloys},
  author={Divilov, Simon and Eckert, Hagen and Toher, Cormac and Friedrich, Rico and Zettel, Adam C and Brenner, Donald W and Fahrenholtz, William G and Wolfe, Douglas E and Zurek, Eva and Maria, Jon-Paul and others},
  journal={Acta Materialia},
  volume={266},
  pages={119667},
  year={2024},
}

@article{rao2021beyond,
  title={Beyond solid solution high-entropy alloys: tailoring magnetic properties via spinodal decomposition},
  author={Rao, Ziyuan and Dutta, Biswanath and K{\"o}rmann, Fritz and Lu, Wenjun and Zhou, Xuyang and Liu, Chang and da Silva, Alisson Kwiatkowski and Wiedwald, Ulf and Spasova, Marina and Farle, Michael and others},
  journal={Advanced Functional Materials},
  volume={31},
  number={7},
  pages={2007668},
  year={2021},
}

@article{cao2009pandat,
  title={{PANDAT software with PanEngine, PanOptimizer and PanPrecipitation for multi-component phase diagram calculation and materials property simulation}},
  author={Cao, Weisheng and Chen, S-L and Zhang, Fan and Wu, K and Yang, Y and Chang, YA and Schmid-Fetzer, R and Oates, WA},
  journal={Calphad},
  volume={33},
  number={2},
  pages={328--342},
  year={2009},
}

@misc{sariturk_2025_15730911,
  author={Sar{\i}t{\"u}rk, Do{\u{g}}uhan and Zhu, Siya and Arr{\'o}yave, Raymundo},
  title={{PhaseForge}},
  month=jun,
  year=2025,
  publisher={Zenodo},
  doi={10.5281/zenodo.15730911},
}

@misc{sariturk_2025_15731044,
  author={Sar{\i}t{\"u}rk, Do{\u{g}}uhan and Arr{\'o}yave, Raymundo},
  title={{MaterialsFramework}},
  month=jun,
  year=2025,
  publisher={Zenodo},
  doi={10.5281/zenodo.15731044},
}

@article{vela2025visualizing,
  title={Visualizing high entropy alloy spaces: methods and best practices},
  author={Vela, Brent and Hastings, Trevor and Allen, Marshall and Arr{\'o}yave, Raymundo},
  journal={Digital Discovery},
  volume={4},
  number={1},
  pages={181--194},
  year={2025},
}

@article{kunselman2025construction,
  title={{Construction and Tuning of CALPHAD Models Using Machine-Learned Interatomic Potentials and Experimental Data: A Case Study of the Pt--W System}},
  author={Kunselman, Courtney and Zhu, Siya and Sar{\i}t{\"u}rk, Do{\u{g}}uhan and Arr{\'o}yave, Raymundo},
  journal={Journal of Phase Equilibria and Diffusion},
  pages={1--9},
  year={2025},
}

@article{bocklund2019espei,
  title={{ESPEI for efficient thermodynamic database development, modification, and uncertainty quantification: application to Cu--Mg}},
  author={Bocklund, Brandon and Otis, Richard and Egorov, Aleksei and Obaied, Abdulmonem and Roslyakova, Irina and Liu, Zi-Kui},
  journal={MRS Communications},
  volume={9},
  number={2},
  pages={618--627},
  year={2019},
}

@article{de1973analysis,
  title={{An analysis of clustering and ordering in multicomponent solid solutions—II fluctuations and kinetics}},
  author={De Fontaine, D},
  journal={Journal of Physics and Chemistry of Solids},
  volume={34},
  number={8},
  pages={1285--1304},
  year={1973},
}

@article{tandoc2025bond,
  title={A bond-based model for accurate prediction of lattice parameters of bcc solid solution alloys},
  author={Tandoc, Christopher and Qi, Liang and Hu, Yong-Jie},
  journal={Materialia},
  volume={40},
  pages={102410},
  year={2025},
}

@article{zhu2025accelerating,
  title={{Accelerating CALPHAD-based phase diagram predictions in complex alloys using universal machine learning potentials: Opportunities and challenges}},
  author={Zhu, Siya and Sar{\i}t{\"u}rk, Do{\u{g}}uhan and Arr{\'o}yave, Raymundo},
  journal={Acta Materialia},
  volume={286},
  pages={120747},
  year={2025},
}

@article{zhu2025machine,
  title={{Machine learning potentials for alloys: a detailed workflow to predict phase diagrams and benchmark accuracy}},
  author={Zhu, Siya and Sar{\i}t{\"u}rk, Do{\u{g}}uhan and Arr{\'o}yave, Raymundo},
  journal={npj Computational Materials},
  volume={11},
  pages={340},
  year={2025},
}

@article{zunger1990special,
  title={Special quasirandom structures},
  author={Zunger, Alex and Wei, Su-Huai and Ferreira, L. G. and Bernard, James E.},
  journal={Physical Review Letters},
  volume={65},
  number={3},
  pages={353--356},
  year={1990},
}

@article{van2002alloy,
  title={{The alloy theoretic automated toolkit: A user guide}},
  author={van de Walle, Axel and Asta, Mark and Ceder, Gerbrand},
  journal={Calphad},
  volume={26},
  number={4},
  pages={539--553},
  year={2002},
}

@article{dinsdale1991sgte,
  title={{SGTE} data for pure elements},
  author={Dinsdale, A. T.},
  journal={Calphad},
  volume={15},
  number={4},
  pages={317--425},
  year={1991},
}

@misc{neumann2024orb,
  title={{Orb: A Fast, Scalable Neural Network Potential}},
  author={Neumann, Mark and Gin, James and Rhodes, Benjamin and Bennett, Steven and Li, Zhiyi and Choubisa, Hitarth and Hussey, Arthur and Godwin, Jonathan},
  year={2024},
  eprint={2410.22570},
  archivePrefix={arXiv},
  primaryClass={cond-mat.mtrl-sci}
}

@misc{deng2023chgnet,
  title={{CHGNet: Pretrained Universal Neural Network Potential for Charge-Informed Atomistic Modeling}},
  author={Deng, Bowen and Zhong, Peichen and Jun, KyuJung and Riebesell, Janosh and Han, Kevin and Bartel, Christopher J. and Ceder, Gerbrand},
  year={2023},
  eprint={2302.14231},
  archivePrefix={arXiv},
  primaryClass={cond-mat.mtrl-sci}
}

@article{schmidt2023alexandria,
  title={{Machine-Learning-Assisted Determination of the Global Zero-Temperature Phase Diagram of Materials}},
  author={Schmidt, Jonathan and Hoffmann, Noah and Wang, Hai-Chen and Borlido, Pedro and Carri{\c{c}}o, Pedro J. M. A. and Cerqueira, Tiago F. T. and Botti, Silvana and Marques, Miguel A. L.},
  journal={Advanced Materials},
  year={2023},
  doi={10.1002/adma.202210788}
}

@article{cahn1962coherent,
  title={Coherent fluctuations and nucleation in isotropic solids},
  author={Cahn, John W},
  journal={Acta Metallurgica},
  volume={10},
  number={10},
  pages={907--913},
  year={1962},
  publisher={Elsevier}
}

@article{virtanen2020scipy,
  title={SciPy 1.0: fundamental algorithms for scientific computing in Python},
  author={Virtanen, Pauli and Gommers, Ralf and Oliphant, Travis E and Haberland, Matt and Reddy, Tyler and Cournapeau, David and Burovski, Evgeni and Peterson, Pearu and Weckesser, Warren and Bright, Jonathan and others},
  journal={Nature methods},
  volume={17},
  number={3},
  pages={261--272},
  year={2020},
  publisher={Nature Publishing Group US New York}
}

@article{cahn1958free,
  title={Free energy of a nonuniform system. I. Interfacial free energy},
  author={Cahn, John W and Hilliard, John E},
  journal={The Journal of chemical physics},
  volume={28},
  number={2},
  pages={258--267},
  year={1958},
  publisher={American Institute of Physics}
}

@article{eyre1998unconditionally,
  title={Unconditionally gradient stable time marching the Cahn-Hilliard equation},
  author={Eyre, David J},
  journal={MRS online proceedings library (OPL)},
  volume={529},
  pages={39},
  year={1998},
  publisher={Cambridge University Press}
}

@article{shen2010numerical,
  title={Numerical approximations of allen-cahn and cahn-hilliard equations},
  author={Shen, Jie and Yang, Xiaofeng},
  journal={Discrete Contin. Dyn. Syst},
  volume={28},
  number={4},
  pages={1669--1691},
  year={2010}
}

@article{moulinec1998numerical,
  title={A numerical method for computing the overall response of nonlinear composites with complex microstructure},
  author={Moulinec, Herv{\'e} and Suquet, Pierre},
  journal={Computer methods in applied mechanics and engineering},
  volume={157},
  number={1-2},
  pages={69--94},
  year={1998},
  publisher={Elsevier}
}

@article{chen1998applications,
  title={Applications of semi-implicit Fourier-spectral method to phase field equations},
  author={Chen, Long Qing and Shen, Jie},
  journal={Computer Physics Communications},
  volume={108},
  number={2-3},
  pages={147--158},
  year={1998},
  publisher={Elsevier}
}

@inproceedings{clough1965finite,
  title={Finite element stiffness matricess for analysis of plate bending},
  author={Clough, Ray W},
  booktitle={Proc. of the First Conf. on Matrix Methods in Struct. Mech.},
  pages={515--546},
  year={1965}
}

@article{gururajan2007phase,
  title={Phase field study of precipitate rafting under a uniaxial stress},
  author={Gururajan, MP and Abinandanan, TA},
  journal={Acta Materialia},
  volume={55},
  number={15},
  pages={5015--5026},
  year={2007},
  publisher={Elsevier}
}

@article{de1979configurational,
  title={Configurational thermodynamics of solid solutions},
  author={De Fontaine, Didier},
  journal={Solid state physics},
  volume={34},
  pages={73--274},
  year={1979},
  publisher={Elsevier}
}

@book{pearson2013handbook,
  title={A handbook of lattice spacings and structures of metals and alloys: International series of monographs on metal physics and physical metallurgy, Vol. 4},
  author={Pearson, William Burton},
  volume={4},
  year={2013},
  publisher={Elsevier}
}

@article{de2015charting,
  title={Charting the complete elastic properties of inorganic crystalline compounds},
  author={De Jong, Maarten and Chen, Wei and Angsten, Thomas and Jain, Anubhav and Notestine, Randy and Gamst, Anthony and Sluiter, Marcel and Krishna Ande, Chaitanya and Van Der Zwaag, Sybrand and Plata, Jose J and others},
  journal={Scientific data},
  volume={2},
  number={1},
  pages={150009},
  year={2015},
  publisher={Nature Publishing Group}
}

@article{soderlind1993theory,
  title={Theory of elastic constants of cubic transition metals and alloys},
  author={S{\"o}derlind, Per and Eriksson, Olle and Wills, JM and Boring, AM},
  journal={Physical Review B},
  volume={48},
  number={9},
  pages={5844},
  year={1993},
  publisher={APS}
}

@article{ikeda2019ab,
  title={Ab initio phase stabilities and mechanical properties of multicomponent alloys: A comprehensive review for high entropy alloys and compositionally complex alloys},
  author={Ikeda, Yuji and Grabowski, Blazej and K{\"o}rmann, Fritz},
  journal={Materials Characterization},
  volume={147},
  pages={464--511},
  year={2019},
  publisher={Elsevier}
}

\end{document}